\documentclass[fleqn,usenatbib]{mnras}

\usepackage[T1]{fontenc}

\DeclareRobustCommand{\VAN}[3]{#2}
\let\VANthebibliography\thebibliography
\def\thebibliography{\DeclareRobustCommand{\VAN}[3]{##3}\VANthebibliography}
\usepackage{graphicx}	% Including figure files
\usepackage{amsmath}	% Advanced maths commands
\usepackage{amssymb}	% Extra maths symbols
\usepackage{subfigure}
\usepackage[labelfont=bf,textfont=it,labelsep=endash]{subcaption} % ugly setting, just for demonstration
\usepackage{stackengine}
\usepackage{newtxtext,newtxmath}
\makeatletter
\newcommand\makesimplesubcaption{\caption@@@make{\caption@fnum{sub\@captype}}}
\makeatother

\title[Black hole masses in barred and unbarred IllustrisTNG galaxies]{How does the presence of bar affects the fueling of supermassive black holes ? An IllustrisTNG100 perspective}

\author[Kataria $\&$  Vivek]{
Sandeep Kumar Kataria $^{1,2}$\thanks{skkataria.iit@gmail.com} \&
 M. Vivek,$^{3}$\thanks{vivek.m@iiap.res.in} 
\\
$^{1}$Department of Astronomy, School of Physics and Astronomy, Shanghai Jiao Tong University, 800 Dongchuan Road, Shanghai 200240, China\\
$^{2}$Key Laboratory for Particle Astrophysics and Cosmology (MOE) / Shanghai Key Laboratory for Particle Physics and Cosmology, Shanghai 200240, China\\
$^{3}$Indian Institute of Astrophysics, Bangalore 560034, India \\
}

\date{Accepted XXX. Received YYY; in original form ZZZ}

\pubyear{2015}

\begin{document}
\label{firstpage}
\pagerange{\pageref{firstpage}--\pageref{lastpage}}
\maketitle

\begin{abstract}

We conduct a statistical study of black hole masses of barred and unbarred galaxies in the IllustrisTNG100 cosmological magneto-hydrodynamical simulations. This work aims to understand the role of the bars in the growth of central supermassive black hole mass and its implications on AGN fueling. Our sample consists of 1191 barred galaxies and 2738 unbarred galaxies in the IllustrisTNG100 simulations. To have an unbiased study, we perform our analysis with an equal number of barred and unbarred galaxies by using various controlled parameters like total galaxy mass, stellar mass, gas mass, dark matter halo mass, etc. Except for the stellar mass controlling, we find that the median of the black hole mass distribution for barred galaxies is higher than that of the unbarred ones indicating that stellar mass is a key parameter influencing the black hole growth. The higher mean accretion rate of the black holes in barred galaxies, averaged since the bar forming epoch (z$\sim$2 ), explains the higher mean black hole masses in barred galaxies. Further, we also test that these results are unaffected by other environmental processes like minor/major merger histories and neighboring gas density of black hole. Although the relationship between stellar mass, bar formation, and black hole growth is complex, with various mechanisms involved, our analysis suggests that bars can play a crucial role in feeding black holes, particularly in galaxies with massive stellar disks.

\end{abstract}


%%%%%%%%%%%%%%%%%%%%%%%%%%%%%%%%%%%%%%%%%%%%%%%%%%

%%%%%%%%%%%%%%%%% BODY OF PAPER %%%%%%%%%%%%%%%%%%

\section{Introduction}
{ An Active Galactic Nucleus (AGN)} plays a vital role in galaxy evolution by providing feedback to the host galaxy \citep{Silk.Rees.1998}. In this feedback, AGNs release part of their rest mass accreted energy to the gas reservoir in their host galaxy \citep{Fabian.2012}. AGN and stellar feedback are considered necessary in high and low-mass galaxies, respectively, which explains the observed galaxy luminosity function. Many semi-analytical and numerical simulations have shown the importance of feedback \citep[for e.g.,][]{Kauffmann.Haehnnelt.et.al.2000,Granato.et.al.2004,Springel.2010}.
It has been observed that the most massive galaxies tend to have the most massive SMBHs at their centers \citep{Kormendy.Ho.2013}. Additionally, there is a correlation between the mass of the SMBH and the properties of the host galaxy, such as the bulge mass and velocity dispersion \citep{Ferrarese2000}. These observations suggest that the growth of SMBHs and the feedback they provide play a crucial role in shaping the properties of galaxies.
% \citet{Kormendy.Ho.2013} shows that 2\% of ellipticals, 5\% of classical bulges, and 50\% of pseudo bulges possess Seyfert nuclei ({\color{cyan} VM: tried to find proof of this statement in Kormendy \& Ho paper, But couldn't find}). Here, the classical bulges correspond to dispersion-dominated elliptical galaxy systems while pseudo bulges correspond to rotation-dominated bulges, which result from the secular evolution of disk galaxies \citep{Laurikainen.Salo.2016}. These observed statistics suggest that active nuclei are typically present in disk-type galaxies.  

The disk galaxies in the observable universe come in various shapes and sizes \citep{Binney.Tremaine.2008}. {It is well known that non-axisymmetric instabilities, e.g., bar and spirals in the disk galaxy, play an essential role in the secular evolution of disk galaxies, leading to the formation of pseudo bulges whereas classical bulges are formed through merger events \citep{Kormendy.Kennicutt.2004}.  Many N-body simulation studies show the formation of Box/Peanut/X-shaped bulges in barred galaxies during their secular evolution via the bending of bar or 2:1 vertical resonances. \citep{Raha.et.al.1991,Pfenniger.Friedli.1991,Shen.et.al.2010, Zana2019,Lokas2019}. { As the bar forms and rotates, it exerts torques on the surrounding material. This torque extracts angular momentum from the inner regions, causing material to move towards the central region.} Thus, the bar plays a significant role in transporting angular momentum from the central region to the outer parts of disks, which leads to the transport of material within the disk \citep{Athanassoula.2003}.  }%It is well known that non-axisymmetric instabilities, e.g. bar and spirals in the disk galaxy, play an essential role in the secular evolution of disk galaxies, leading to the formation of pseudo bulges \citep{Kormendy.Kennicutt.2004}. The bar plays a significant role in transporting angular momentum from the central region to the outer parts of disks \citep{Athanassoula.2003}. Further, many N-body simulation studies show the formation of Box/Peanut/X-shaped bulges in barred galaxies during its secular evolution \citep{Raha.et.al.1991,Pfenniger.Friedli.1991,Shen.et.al.2010} 
%{\color{cyan} VM: This is essentially just a rearrangement of previous text. This is not reading well.}.  

{ Observations claim that nearly two third of massive disk galaxies possess a bar in the disks, as evidenced by optical and infrared surveys  \citep{Eskridge.et.al.2000, Delmestre.et.al.2007,Erwin.2018}. It is understood that the fraction of barred galaxies is contingent upon the properties of the host galaxies like mass, as well as the methods employed for identifying bars within them \citep{Lee2019}.} Bars are non-axisymmetric (m=2 Fourier mode) instabilities that form in kinematically colder disks of galaxies \citep{Ostriker.Peebles.1973}. Recent advances in observational technology, such as the high resolving power of the James Webb Space Telescope (JWST), have enabled the identification of bars at high redshifts (z>1) \citep{Guo.et.al.2023}. The fraction of barred galaxies is constant up to z $\sim$ 0.84 for massive spiral galaxies while the bar fraction decline beyond z $\sim$ 0.3 for low mass spiral galaxies \citep{Sheth.et.al.2008}. { This suggests that the bar can survive in  disk galaxies for a long time, and its evolution can play a significant role in disk dynamics.}  The presence of a central massive concentration like bulge is shown to be destructive for bar survival in the disk \citep{Nelson.et.al.2018,Shen.Sellwood.2004,Kataria.Das.2018,Kataria.Das.2019,Kataria2020,Jang.Kim.2022}. The surrounding dark matter halo properties also play an essential role in the formation and evolution of bars and central X-shaped bulges mostly seen in barred galaxies \citep{Kanak.Saha.Naab.2013,Longetal.2014,Collieretal.2018,Collieretal.2019,Kumar.et.al.2022,Kataria.Shen.2022,Ansar.et.al.2023}. Recently large-scale magneto-hydrodynamical cosmological simulations have started producing bars whose properties are qualitatively similar to the observed ones though relatively smaller in sizes \citep{Rosas-Guevara.et.al.2020,Zhao.et.al.2020,Frankel.et.al.2022,Rosas2022}.

Bars are essential in transporting gas from the outer disk to the central disk \citep{Sakamoto.et.al.1999}. Large-scale bars, also referred to as primary bars, can transfer gas up to the central kpc region. However, the local viscous stresses responsible for gas transport \citep{Balbus.Hawley.1991} becomes inefficient within the sub-parsec region of disk \citep{Sholsman.et.al.1989,Shosman.et.al.1990}. In a theoretical study, \cite{Sholsman.et.al.1989}  suggested ``bars within bars" as a possible mechanism to transport gas inwards in the central sub-parsec region. This mechanism is motivated by the required fueling of gas in a given time that can sustain AGN activity. {At galactic scales, the primary bars transport gas to the inner $\sim$ 1 kpc, where gas accumulates. Further, the gas is sequentially transported inwards when the accumulated gas fraction becomes high at these scales and becomes susceptible to gravitational instabilities due to self-gravitation. These instabilities form structures like nuclear spirals, bars, rings, and clumpy disks that trigger nuclear star formation and gas inflows. These nuclear instabilities transport gas efficiently to the inner regions, where another set of instabilities develop and transport gas further inwards. } Using simulations, \cite{Hopkins.Quataert.2010} showed that structures like spirals, rings, bars, etc., in the nuclear regions are efficient in transporting gas within the central sub-parsec regions. 
 %using around 100 simulations to have ample parameters space of galaxy properties and various models of the interstellar medium.
 These simulations contained various models of the interstellar medium that properly treat star formation and self-gravity of stars and gas, which are crucial for exploring the central sub-parsec region. This simulation study nicely supplemented the ``bars within bars" proposition of \cite{Sholsman.et.al.1989}.     
 
 {Observationally, the approach to study the connection between bars and AGN activity has been twofold: One approach compared bar fraction in a sample of AGN host galaxies with that of a control sample of inactive galaxies, and the second approach searched for differences in AGN activity in barred and unbarred galaxies \citep[See][ for a review]{Storchi2019}. In the former approach, three studies, \citet{Knapen.et.al.2000} (79\% vs. 59\%), \citet{Laine.et.al.2002} (73\% vs. 50\%)  and \citet{Garland.et.al.2023} (59\% vs. 44 \%) found that Seyfert galaxies are found to have more bars than non-active galaxies. \citep{Laurikainen.et.al.2004} also found a similar result for SB-type bars in their analysis. \citet{Galloway.et.al.2015} created a volume-limited sample of active and inactive disk galaxies using the Galaxy Zoo 2 project and found a small but statistically significant increase in the bar fraction of active galaxies. However, \citet{Ho.et.al.1997,Mulchaey.Regan.1997,Lee_2012} detected similar bar fractions for Seyfert galaxies and comparison galaxies. Following the alternate approach,
 \citet{Alonso2013} and \citet{Alonso.et.al.2018} compared the nuclear activity of barred galaxies with that of unbarred galaxies and AGN galaxies in paired systems, respectively. In both studies,  barred galaxies showed excess nuclear activity and black hole accretion rate compared to the control samples. However, \citet{Cisternas.et.al.2013} did not find any significant correlation between any of the bar strength indicators and the degree of nuclear activity.
 Similarly, \citep{Goulding.et.al.2017} used Galaxy Zoo 2 project and Chandra X-ray data and showed no enhancement in the nuclear accretion with the presence of a bar in the host galaxy. \citet{Cheung2015} combined the two approaches by comparing a sample of 120 AGN host galaxies in the redshift range, 0.2 $<$ z $<$ 1.0, with a control sample of inactive galaxies. They found no significant difference in the bar fraction among active and inactive galaxies and AGN fraction among barred and non-barred galaxies. Finally, the most recent study by \citet{Silva-Lima.et.al.2022}  points out that AGNs are favored in barred galaxies though it also suggests other mechanisms within the 100 pc region that can feed gas to AGNs. Clearly, there is no consensus among the observational studies about the role of bars in AGN fueling.
 
 }
 
 %In literature, several observational studies have probed whether bar in the host galaxy affects AGN activity or not. Some observational studies with optical and near-infrared morphological images of \cite{Ho.et.al.1997,Mulchaey.Regan.1997} show no correlation between AGN activity and the presence of a bar. 
 %A recent study  using  SDSS Galaxy Zoo 2 Project and Chandra X-ray data show that there is no enhancement in the nuclear accretion with the presence of bar in the host galaxy \citep{Goulding.et.al.2017}. 
% Whereas, another set of studies using  a volume-limited SDSS Galaxy Zoo 2  sample and near-infrared data show  enhanced presence of bars in AGNs compared to unbarred galaxies \citep{Knapen.et.al.2000,Galloway.et.al.2015}. A recent study has identified bars using 2D disk decomposition by BUDDA  and emission line diagnostics to identify AGN activity\citep{Silva-Lima.et.al.2022}. This study points out that AGNs are favored in barred galaxies though it also suggests other mechanisms within the 100 pc region that can feed gas to AGNs. The bar and AGN connection still need to be clarified, given the conflicting results from observational studies. 

{ The new generation of cosmological hydrodynamical simulations offers a powerful tool to study the role of various galaxy components in the evolution of galaxies \citep{Vogelsberger.et.al.2014a}.  { The suite of these simulations include EAGLE \citep{Schaye.et.al.2015}, ErisBH \citep{Bonoli.Silvia.et.al.2016},  Illustris \citep{Vogelsberger.et.al.2014}, IllustrisTNG \citep{Nelson.et.al.2018} and MassiveBlack-II \citep{Khandai.et.al.2015}}. These simulations are able to correctly reproduce many observed properties of galaxies at low redshift. Among these, the IllustrisTNG project, the successor of the original Illustris simulations, contains the most recent physical models and an updated prescription for feedback processes that enable the generation of galaxy populations whose properties are even closer to reality. There are several studies that attempted to study the bars in galaxies using these simulations. \citet{Peschken2019} studied the bar formation in galaxies using Illustris simulation and suggested that a large fraction of bars are formed by mergers and flyby events. { \citet{Zana2018a} demonstrated that the global properties of bar formation remained unaffected by mergers and flybys in a variant of the ErisBH simulation. Furthermore, in the context of Eris2k simulations, \citet{Zana2018b} established that bars are resilient to flyby interactions.} Using the Fourier decomposition method, \citet{Rosas-Guevara.et.al.2020} showed that bars are found in 40 \% of the IllustrisTNG disk galaxies identified at z=0 in the stellar mass range 10$^{10.4-11}$ M$_{\odot}$. Their study concludes that the cosmological environment and feedback processes determine the chances of forming bars in a galaxy. A similar result is obtained by \citet{Zhou2020} in their comparison study on the bar structure in the Illustris and IllustrisTNG100 (hereafter TNG100) simulations. However, \citet{Zhao.et.al.2020} revisited the bar fraction using the ellipse fitting method and reported that 55\% of disk galaxies with stellar mass M$_{\star}$ $\approx$ 10$^{10.6}$ M$_{\odot}$ are barred, consistent with observations. { In the high-resolution IllustrisTNG50 simulation, \citet{Zana2022} demonstrated that massive galaxies have a bar fraction of 60\% at z=0, which is comparable to observed values.}%{\color{cyan} VM: Include any other Illustris based bar studies that I may have forgotten to include.}

Several observational and theoretical studies suggest that AGN have a relatively short lifetime, ranging from a few 10$^7$--10$^8$ years \citep{Martini2004,wada2004,Merloni2004}. In addition, \citet{wada2004} suggest that mass accretion onto black holes is not constant even during the expected duty cycle of 10$^8$ years, but rather consists of multiple shorter episodes lasting 10$^{4-5}$ years. On the other hand, numerical simulations have shown that bars have a lifetime of around a few Giga years \citep{Rosas-Guevara.et.al.2020}. This difference in the timescales may explain the lack of a clear correlation between the presence of large-scale bars and AGN activity. { Nevertheless, the growth of black holes and AGN activity is the most obvious consequence of gas accretion onto black holes. So, it may be possible to investigate the connection between bars and AGN activity by examining the distribution of black hole masses in these galaxies. However, this is challenging observationally as the number of AGN with robust black hole mass measurements remains tiny. Large-volume cosmological simulations like IllustrisTNG offers a  promising alternative as they include a black hole prescription in the simulation and can resolve morphological features such as bars.
}

{In this work, we use the IllustrisTNG simulation to study the role of bars in the growth of central supermassive black holes.
We use the  TNG100 simulation, which has a nice compromise between a large cosmological volume and resolution.}
The paper is structured as follows. In section \ref{methods}, we discuss the methods and techniques used in this study. We discuss the results and discussion in sections \ref{results} and \ref{Discussion}, respectively. Finally, we conclude our results in section \ref{conclusion}. 

%This article gives insights into the following questions in TNG100 simulations.
%Questions 1) Does the presence of a bar affect the growth of black hole masses in the host galaxy? 2) How does the accretion rate change with the presence of bars in the disk of the host galaxy? 

\section{Methods} \label{methods}
\subsection{IllustrisTNG Simulations}\label{simulations}
IllustrisTNG simulations are the gravo-magnetohydrodynamical cosmological simulations \citep{Marinacci.et.al.2018,Naiman.et.al.2018,Nelson.et.al.2018,Pillepich.et.al.2018,Pillepich.et.al.2019,Springel.et.al.2018} which are run with moving mesh code AREPO \citep{Springel.2010,Pakmor.et.al.2016}.  IllustrisTNG simulations are the successor to Illustris simulations \citep{Genel.et.al.2014,Vogelsberger.et.al.2014a,Vogelsberger.et.al.2014b,Sijacki.et.al.2015} which incorporates new physics and new treatment of cosmic magnetic field \citep{Weinberger.et.al.2017,Pillepich.et.al.2018b} and these simulations solve magnetohdrodynamics equations using finite volume Godunov type method coupled with self-gravity using Tree-PM methods. 
{ The IllustrisTNG simulations are made of three cosmological volumes, TNG50, TNG100, and TNG300, which have comoving box sizes of roughly 51.7, 110.7, and 302.6 Mpc, respectively.} Among these simulations with increasing box sizes, TNG50 corresponds to the highest resolution, while TNG300 corresponds to the lowest resolution. We have used TNG100 simulations for this study \citep{Nelson.et.al.2018}, which provide a good balance between cosmological volume and resolution of the simulations. The mass resolutions of gas and dark matter in TNG100 simulations are given by 1.4$\times 10^6$ $M_\odot$ and 7.5$\times 10^6$ $M_\odot$, respectively. { The comoving gravitational softening lengths ($\epsilon$) of the gas and dark matter particles are equal to 180 pc and 740 pc, respectively.} These simulations incorporate the cosmology from the \cite{Planck.Collaboration.et.al.2016} for which the parameters are given by $\Omega_\Lambda$ = 0.6911, $\Omega_m$ = 0.3089, $\Omega_b$ = 0.0486, $H_0$ = 67.74 Km sec$^{-1}$ Mpc$^{-1}$, $\sigma_8$ = 0.8159, n$_s$ = 0.9667. 

In the TNG100  simulation, the black holes are seeded in massive halos having masses $>$ 7.8 $\times$ 10$^{10}$ M$_{\odot}$. The masses of seeded black holes are 1.18 $\times$ 10$^6$ M$_{\odot}$. Supermassive black holes in these simulations can grow through two mechanisms: black hole mergers and gas accretion through secular evolution. The black holes accrete gas via pure Bondi accretion rate, limited by the Eddington rate, as follows: \citep{Weinberger.et.al.2017}. 

\begin{equation}
    \dot{M}_{BH}=min\big(\dot{M}_{Bondi},\dot{M}_{Edd}\big)
\end{equation}

where, 

\begin{equation}
    \centering
    \dot{M}_{Bondi}=\dfrac{4 \pi G^2M_{BH}^2\rho}{c_s^3}
    \label{M_dot}
\end{equation}

\begin{equation}
\centering
    \dot{M_{Edd}}=\dfrac{4 \pi GM_{BH}m_p}{\epsilon_r \sigma_{T} c}
    \label{M_Edd}
\end{equation}

Here, G, M$_{BH}$, $c_s$, and $\rho$ in equation \ref{M_dot} are the Gravitational constant, black hole mass, sound speed, and density, respectively. In equation \ref{M_Edd}, m$_p$, $\sigma_T$, $\epsilon_r$ and $c$ are the mass of the proton, the Thompson cross-section, the radiative accretion efficiency, and speed of light in vacuum, respectively. 
The feedback from the black hole is introduced in two modes: kinetic and thermal modes. The thermal mode feedback is triggered in the high accretion state and heats the surrounding gas with continuous thermal energy. The kinetic mode feedback happens in the low accretion state and emits energy through kinetic outflows that give momentum to the surrounding gas. The kinetic mode feedback is efficient and releases energy in a pulsed manner. The energies released in these two modes are as follows \citep{Weinberger.et.al.2017}.
\begin{equation}
    \Delta E_{thermal}=\epsilon_{f,high} \epsilon_{r} \dot{M}_{BH} c^2
\end{equation}

\begin{equation}
    \Delta E_{kinetic}=\epsilon_{f,low} \dot{M}_{BH} c^2
\end{equation}

Here, $\dot{M}_{BH}$ is the accretion rate of the black hole with mass $M_{BH}$, $\epsilon_r$ is the radiative efficiency typically ranging from 0.1 to 0.2, $\epsilon_{f,high}$ and $\epsilon_{f,low}$ are the fraction of energy released to the surrounding gas for thermal and kinematic mode, respectively. In high accretion rate, the energy is liberated at a constant rate, i.e., $\epsilon_{f,high}=0.1$ ; $\epsilon_{f,high} \epsilon_{r}=0.02$. The maximum limit on emitted fractional energy for low accretion rate mode is $\epsilon_{f,low}=0.2$. 
{In TNG100 simulations, for black hole masses less than $10^{8.5}$,  thermal feedback is the main feedback mechanism, and the black holes are in the high accretion state. In massive black holes, $M_{BH} >10^{8.5}$, gas accretion is self-regulated via the more efficient kinetic feedback, which causes their Eddington ratios to drop \citep{Weinberger.et.al.2018}.  }

% \begin{figure*}
%    \begin{tabular}{c|c|c|c}
%     ID=102685  \hspace{2.6 cm}  ID=102689 \hspace{2 cm}     ID=102691  \hspace{2 cm}    ID=102700\\
%     \subfigure{\includegraphics[width=0.20\textwidth]{vis_TNG100-1_subhalo_99_102685_stars_density.png}
%     }   
%     \subfigure{\includegraphics[width=0.20\textwidth]{vis_TNG100-1_subhalo_99_102689_stars_density.png}} 
%     \subfigure{\includegraphics[width=0.20\textwidth]{vis_TNG100-1_subhalo_99_102691_stars_density.png}} 
%     \subfigure{\includegraphics[width=0.20\textwidth]{vis_TNG100-1_subhalo_99_102700_stars_density.png}} \\
%     ID=20  \hspace{2.8 cm}  ID=32 \hspace{3 cm}     ID=33  \hspace{3 cm}    ID=48\\    
%     \subfigure{\includegraphics[width=0.20\textwidth]{vis_TNG100-1_subhalo_99_20_stars_density.png}}
%     \subfigure{\includegraphics[width=0.20\textwidth]{vis_TNG100-1_subhalo_99_32_stars_density.png}}
%     \subfigure{\includegraphics[width=0.20\textwidth]{vis_TNG100-1_subhalo_99_33_stars_density.png}}
%     \subfigure{\includegraphics[width=0.20\textwidth]{vis_TNG100-1_subhalo_99_48_stars_density.png}}    
%    \end{tabular}

%     \caption{The top row shows some examples of face-on stellar density maps for barred galaxies at z=0 in the TNG100 simulation. The bottom row shows some examples of face-on stellar density maps for unbarred galaxies at z=0 in the TNG100 simulation. The sub-halo ids of all galaxies are mentioned on top of each galaxy.  }
%     \label{fig:Faceon_maps}
% \end{figure*}

\begin{figure*}
   \begin{tabular}{c|c|c|c}
   % ID=102685  \hspace{2.8 cm}  ID=102689 \hspace{2.8 cm}     ID=102691  \hspace{2.8 cm}    ID=102700\\
    \subfigure{\includegraphics[width=0.24\textwidth]{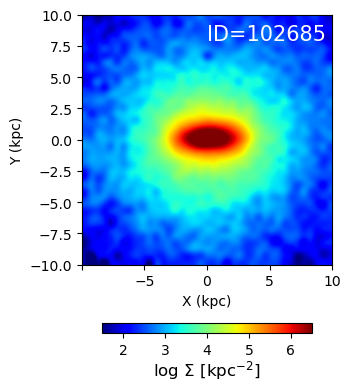}}   
    \subfigure{\includegraphics[width=0.24\textwidth]{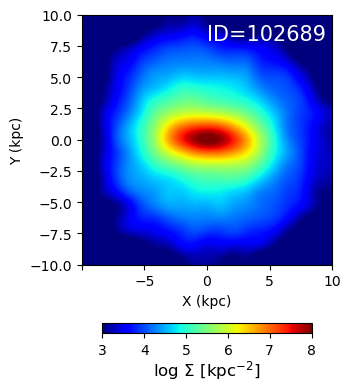}} 
    \subfigure{\includegraphics[width=0.24\textwidth]{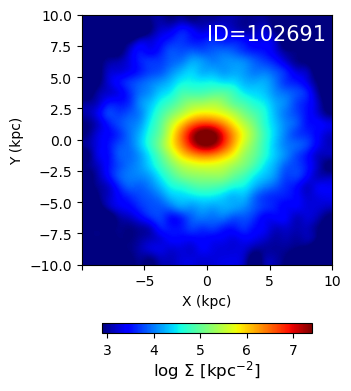}} 
    \subfigure{\includegraphics[width=0.24\textwidth]{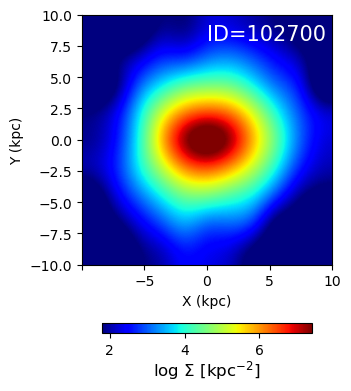}} \\
  %  ID=20  \hspace{3.4 cm}  ID=32 \hspace{3.4 cm}     ID=33  \hspace{3.4 cm}    ID=48\\    
    \subfigure{\includegraphics[width=0.24\textwidth]{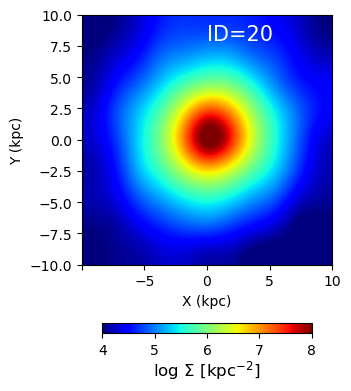}}
    \subfigure{\includegraphics[width=0.24\textwidth]{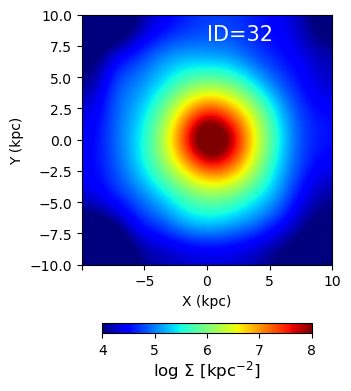}}
    \subfigure{\includegraphics[width=0.24\textwidth]{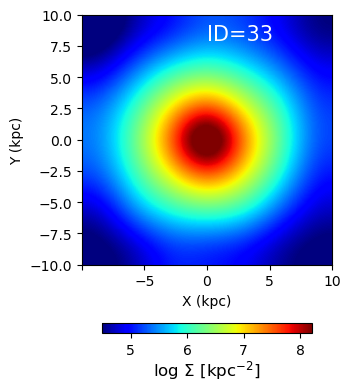}}
    \subfigure{\includegraphics[width=0.24\textwidth]{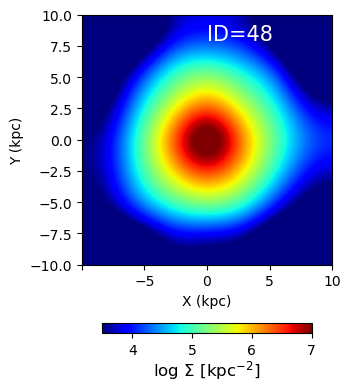}}    
   \end{tabular}

    \caption{{  The top row shows some examples of face-on stellar density maps for barred galaxies at z=0 in the TNG100 simulation. The bottom row shows some examples of face-on stellar density maps for unbarred galaxies at z=0 in the TNG100 simulation. The sub-halo ids of all galaxies are mentioned on top of each galaxy. The color bars within each map correspond to the logarithm of the average stellar volume density in units of kpc$^{-2}$. }}%The color bar scales span the following ranges: ID102685: {-0.9, 0.8}, ID102689: {-1.0, 0.9}, ID102691: {-0.9, 0.7}, ID102700: {-1.8, 0.6}, ID20: {-1.2, 0.7}, ID32: {-1.1, 0.7}, ID33: {-1.1, 0.7}, ID48: {-1.6, 0.7}. }}
    \label{fig:Faceon_maps}
\end{figure*}

\subsection{Galaxy Selection Criterion} \label{selection_criteria}

{ Our sample consists of rotationally dominated disk galaxies from the TNG100 simulation. The rotation dominance is determined using the parameter K$_{rot} =\langle v_{\phi}^2/v^2 \rangle$ \citep{Sales.et.al.2010}, where $v_{\phi}$ and $v$ correspond to the rotational component of velocity and total velocity, respectively. Following \citet{Du.et.al.2019,Du.et.al.2020}, we employed the criterion, K$_{rot}$ $>$ 0.5, to select the sample of rotationally dominated disk galaxies.
%Further, these studies conduct kinematics decomposition \citep{Abadi.et.al.2003} of disk galaxies into several components using an unsupervised machine learning algorithm based on the Gaussian Mixture Model (also called auto-GMM model).
All the disk galaxies have stellar masses greater than 10$^{10}$ M$_{\odot}$, which makes the total sample of disk galaxies equal to 3931. The disk galaxy sample is categorized into barred and unbarred samples using previous studies  \cite{Rosas-Guevara.et.al.2020} and \cite{Zhao.et.al.2020}.}

\cite{Rosas-Guevara.et.al.2020} identified the barred galaxies in the TNG100 simulations using the strength of $m=2$ Fourier mode. Their sample consisted of disk galaxies having stellar masses greater than 10$^{10.4}$ M$_{\odot}$. %The Fourier mode method is most efficient in capturing strong bars in the disk. 
Their study classified the galaxy as a strongly barred galaxy if the strength of $m=2$ Fourier mode is greater than 0.3. The galaxy is termed to have a weak bar if the strength of $m=2$ Fourier mode is between 0.2--0.3. This method captures 107 barred galaxies.

In addition, \cite{Zhao.et.al.2020} used the method of ellipse fitting to identify bars in TNG100 disk galaxies having stellar masses greater than 10$^{10}$ M$_{\odot}$. They measured the ellipticity and position angle of contours by fitting isodensity contours as a function of radius. If the maximum ellipticity of contour at a given radius is higher than 0.4, it is defined as a strong bar and otherwise a weak bar. %Compared to the Fourier mode method, this method's advantage is that it easily captures weak bars. 
\cite{Zhao.et.al.2020} identified  1179 barred galaxies, an order of magnitude larger than detected bars in the Fourier method by \cite{Rosas-Guevara.et.al.2020}. {  The substantial difference in the number of barred galaxies between these studies can be ascribed to the dissimilarity in the initial sample sizes studied. \cite{Rosas-Guevara.et.al.2020} examined galaxies with a stellar mass exceeding 10$^{10.4}$ M$_{\odot}$, while \cite{Zhao.et.al.2020} focused on galaxies with a stellar mass surpassing 10$^{10}$ M$_{\odot}$.  It is worth noting that the larger sample size in \cite{Zhao.et.al.2020} has a lower stellar mass threshold, eliminating concerns regarding incompleteness or missing galaxies between 10$^{10}$ and 10$^{10.4}$ M$_{\odot}$.}

{  After removing common sources from \citet{Rosas-Guevara.et.al.2020}  and \citet{Zhao.et.al.2020} samples, we obtained 1193 barred galaxies.  The sample is the sum of all the barred galaxies coming from the two catalogues, where 93 objects have been identified by both of them.} The remaining sample, i.e.,  2738 out of 3931 disk galaxies, forms the unbarred disk galaxy sample. { From hereon, we use the terms "barred galaxies" and "unbarred galaxies"  to specifically denote galaxies with bars and galaxies without bars, respectively, at a redshift of z=0.} Fig.~\ref{fig:Faceon_maps} shows the face-on stellar density maps of four barred galaxies and four unbarred galaxies in the TNG100 simulation. 

We obtained parameters like blackhole mass, gas mass, stellar mass, total mass and dark matter halo mass, etc., for both barred and unbarred  galaxies.  %We carry out a detailed statistical analysis of these parameters.
In Fig.~\ref{fig:full_sample_dist}, we show the distribution of total galaxy mass, total stellar mass, total gas mass, total stellar by halo mass, total gas by stellar mass, and total gas plus the total stellar mass of both barred and unbarred galaxies. { These distributions show that the median values of various galaxy components are not similar in the full sample of barred and unbarred galaxies. A  Kolmogorov–Smirnov (K–S) test between the barred and unbarred samples returns a p-value close to 1 for different parameters pointing to significant differences in the corresponding distributions. The maximum difference is seen for the stellar mass parameter. Fig.~\ref{fig:full_sample_dist}b shows that barred galaxies have larger stellar mass in comparison to unbarred ones.  } 
%{\color{cyan} (VM: I had specifically asked to discuss Fig.2 here. But, I do not see any discussion. I also see that many of my previous comments are not properly addressed.)}   

\section{Results} \label{results}
In this section, we compare the distributions of various galaxy properties of barred and unbarred galaxy sample.

\begin{figure*}
    \centering
    \includegraphics[scale=.5]{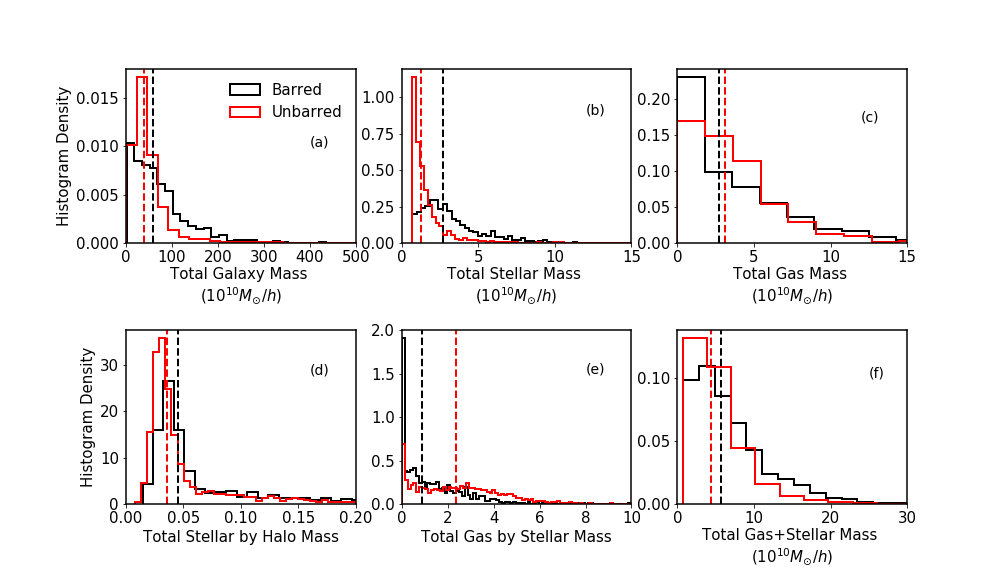}
    \caption{Full sample distribution of various properties of 1193 barred (black histogram) and 2738 unbarred (red histogram) galaxies at z=0, which are used to control the sample; namely (a) total galaxy mass, (b) total stellar mass, (c) total gas mass, (d) total stellar to halo mass ratio, (e) total gas to stellar mass ratio, and (f) sum of gas and stellar mass. Vertical dashed lines show the median values of each distribution. }
    \label{fig:full_sample_dist}
\end{figure*}

\begin{figure}
    \centering
    \includegraphics[scale=0.42]{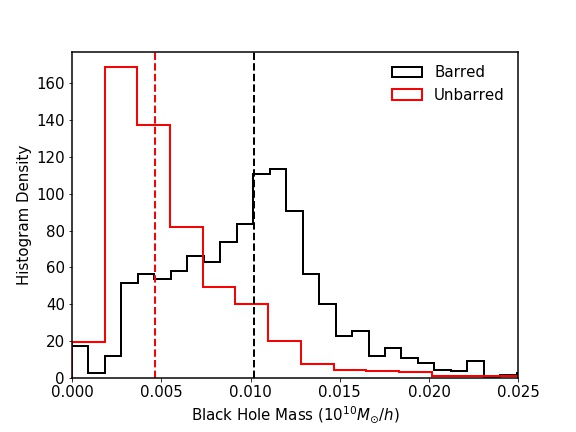}
    \caption{The black hole mass distribution of full barred (black histogram, 1193 galaxies) and unbarred sample (red histogram, 2738 galaxies) at z=0 shows the dichotomy. Vertical dashed lines show the median values of each distribution.}
    \label{fig:Dichotomy_full_sample}
\end{figure}

\begin{figure*}
    \begin{tabular}{cc}
        \includegraphics[scale=0.27]{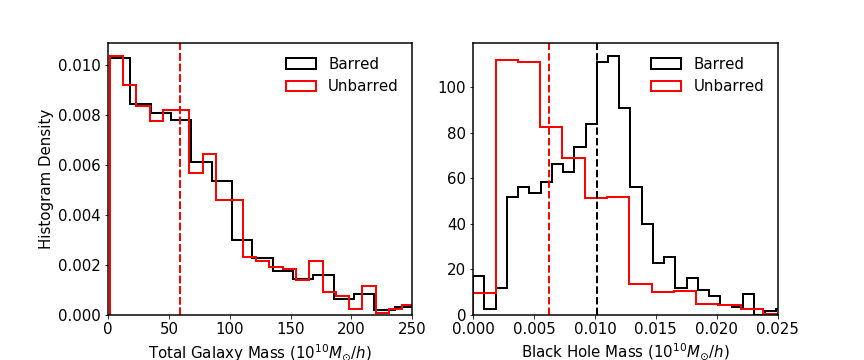}& 
        \includegraphics[scale=0.27]{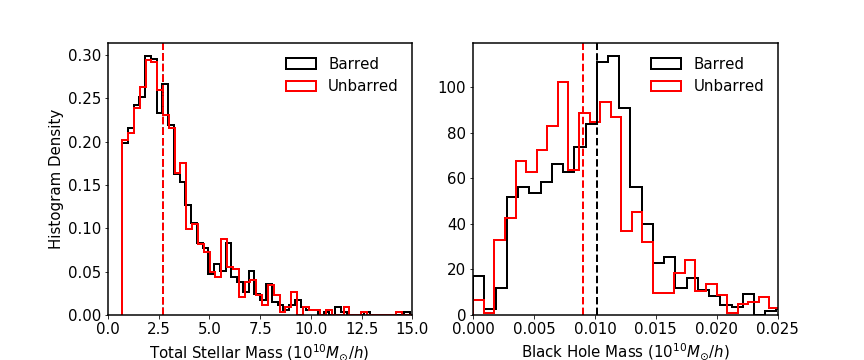} \\
        (a)&(b)\\ 
        \includegraphics[scale=0.27]{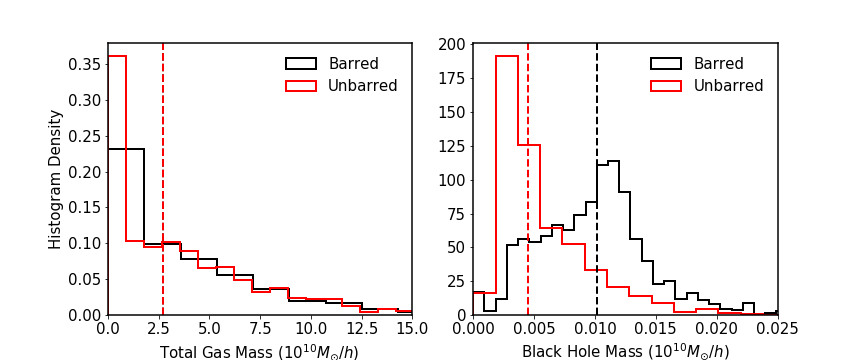}& 
        \includegraphics[scale=0.27]{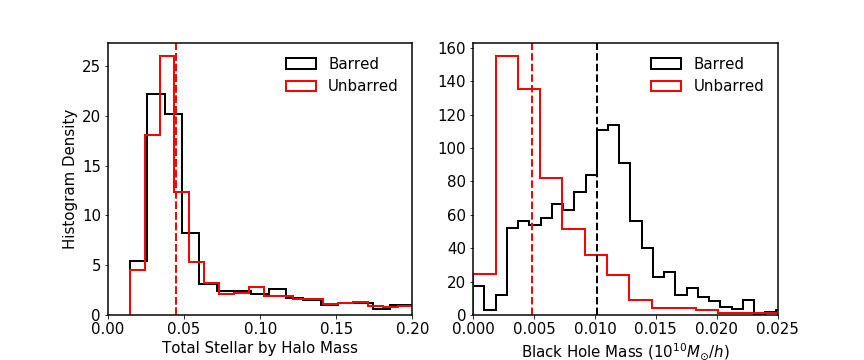} \\
        (c)&(d)\\
        \includegraphics[scale=0.27]{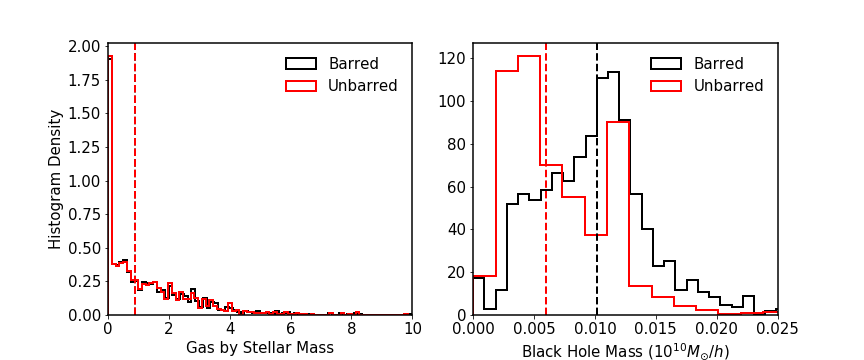}& 
        \includegraphics[scale=0.27]{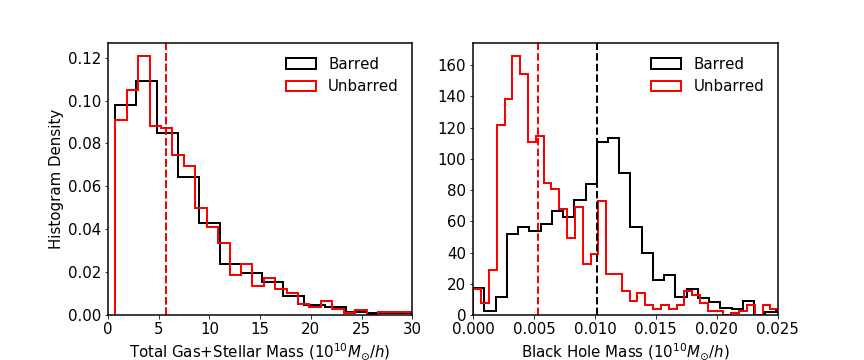}\\
        (e)&(f)

    \end{tabular}
    
    \caption{The distribution of the controlled sample and respective black hole mass distribution of barred (black histogram) \& unbarred (red histogram) galaxies. The left panel of each sub-figure corresponds to the distribution of control parameters; (a) total galaxy mass, (b) total stellar mass, (c) total gas mass, (d) ratio of total stellar and total dark matter halo mass, (e) ratio of total gas mass and total stellar mass (f) sum of total gas mass and total stellar mass. The right panel of each sub-figure corresponds to the distribution of black hole mass at z=0 for the controlled samples. Black/red vertical lines correspond to the median values of the distribution.}
    \label{fig:dichotomy_controlled_sample}
\end{figure*}

\subsection{Black hole mass distribution in barred and unbarred galaxies} \label{BH_dichotomy_barred_unbarred}

As discussed in section \ref{selection_criteria}, we have used 2738 unbarred galaxies and 1193 barred galaxies for this study. As the primary motivation of this study is to investigate the role of galactic bars in fueling the AGN, we first compared the distributions of supermassive black hole mass in these galaxies. Fig.~\ref{fig:Dichotomy_full_sample} shows the black hole mass distribution for the full sample of barred and unbarred galaxies at z=0. These distributions clearly show a mass dichotomy for black hole masses in barred and unbarred samples.  
{ We notice that the peak of the black hole mass distribution for barred galaxies is twice the peak of the black hole mass distribution for unbarred galaxies. { The  difference in the median of black hole mass distributions between barred and unbarred galaxies is $\Delta M_{BH} \approx 4\times10^7 M_{\odot}$ ($\sim$ 42 times the average gas cell mass of 7.93$\times$10$^5$ M$_{\odot}$). } However, the number of unbarred galaxies in our sample is much larger than barred galaxies, which may introduce a bias in our study. To remove any such bias, we made the sample size similar for both barred and unbarred galaxies by creating control samples of unbarred galaxies based on several control parameters. For e.g., in the control sample of unbarred galaxies with stellar mass as the controlled parameter, we only included sources in the unbarred sample with similar stellar mass for each source in the barred sample. This process makes our barred and unbarred sample sizes equal, which is equal to 1193 sources. In our study, we have used several parameters to control the unbarred sample. The control sample generation is implemented using the k-nearest neighbour (KNN) algorithm in the controlled parameter space \citep{Guo2003KNNMA}. { The KNN algorithm is a non-parametric classification and regression technique widely used in data mining and machine learning. In the context of creating a control sample with similar parameter distributions, KNN identifies the most similar data point from a dataset to a given data point in the reference dataset based on their proximity in the feature space. This control sample ensures a balanced comparison between the datasets and minimizes bias.}  The KNN algorithm locates the nearest neighbour in the unbarred sample for each galaxy in the barred sample. We discuss the various controlled parameters and the comparison of black hole mass distributions in the controlled galaxy samples in the following paragraphs. }

% \begin{table*}
%     \centering
%     \begin{tabular}{c|c|c|c}
%         \hline
%          \textbf{Controlled Parameter (CP)} & \textbf{p value of full CP sample} & \textbf{p value of controlled CP sample} & \textbf{p value of black hole mass in controlled sample}    \\
%          \hline
%          Total Galaxy Mass & $10^{-15}$ & 0.99 & $10^{-48}$\\
%          Total Stellar Mass & $10^{-15}$& 0.99 & $10^{-6}$    \\
%          Total Gas Mass & $10^{-13}$ &  0.99 & $0$    \\
%          Total Stellar to Halo Mass & $10^{-41}$& 0.99 & $10^{-15}$\\      
%          Total Gas by Stellar Mass & $10^{-15}$& 0.99 & $10^{-50}$ \\     
%          Total Gas + Stellar Mass & $10^{-15}$& 0.99 & $ 0 $ \\      
%          \hline

%     \end{tabular}
%     \caption{This table shows the p values corresponding to KS test of the distribution of various parameters in barred and unbarred samples used in this study. Column (1) Controlling parameters used to obtain unbiased  barred and unbarred sample (2) p value of KS test for the distribution of controlling parameter in the full barred and unbarred sample (3) p value of the KS test for the distribution of controlling parameter in the controlled sample. (4) p-value of the KS test corresponding to the distribution of black hole masses in controlled barred and unbarred sample. }
%     \label{tab:KS_test}
% \end{table*}

\subsubsection{Sample unbiased using total galaxy mass}
{The total mass of a galaxy plays an essential role in bar formation processes \citep{Athanassoula.2003}. Hence, we first used the total galaxy mass as a control parameter to create unbiased samples of barred and unbarred galaxies. { The total galaxy mass refers to  the total mass associated with the stellar, gaseous, and dark matter components. } The left panel of Fig.~\ref{fig:dichotomy_controlled_sample}(a) shows the distribution of total galaxy masses of barred galaxies and the control sample of unbarred galaxies. The right panel of  Fig.~\ref{fig:dichotomy_controlled_sample}(a) shows the resulting black hole mass distributions of unbiased galaxy samples at z=0. Unlike the galaxy mass distributions for the full sample shown in Fig.~\ref{fig:full_sample_dist}(a), it is clear that the control sample generation has indeed made the total galaxy mass distributions similar for barred and unbarred galaxies. Nevertheless, the black hole mass dichotomy seen in the comparison of the full sample of barred and unbarred galaxies persists. The median value of black hole mass distribution peaks at 0.62 $\times$ 10$^8$ M$_\odot$ for unbarred galaxies and 1.02 $\times$ 10$^8$ M$_\odot$ for barred galaxies. {  The average gas cell mass in the TNG100 simulation is 9.43$\times$10$^5$ M$_{\odot}$  which is much smaller than the difference of median black hole mass in barred and unbarred galaxies.} It should be noted that the halo component makes up the majority of the total mass of a galaxy. { Typically, the halo mass accounts for around 90\% of the total mass of the galaxy, while the stellar and gas components make up 5\% each.} The difference in the distribution of black hole mass between barred and unbarred galaxies, even after controlling by total galaxy mass, suggests that the black hole mass dichotomy is not driven by the difference in the total galaxy mass between the two samples.}
%The distribution of whole galaxy masses for barred and unbarred become exactly similar when the sample is unbiased by choosing the nearest total galaxy mass neighbor of unbarred galaxy corresponding to each barred galaxy as shown in the left column of Fig.~\ref{fig:dichotomy_controlled_sample} (a). 
%The right panel of Fig.~\ref{fig:dichotomy_controlled_sample} (a) shows the black hole mass distribution of barred and unbarred galaxies at z=0 in the controlled sample. We can clearly say from an unbiased sample that there is an apparent dichotomy in the black hole mass distribution of barred and unbarred. The median value of black hole mass distribution peaks at 0.62 x $10^8$ for unbarred and 1.02 x $10^8$ $M_\odot$ for unbarred galaxies. This dichotomy indicates the role of the bar is responsible for the enhanced growth of black hole mass in barred samples compared to unbarred samples.   

\begin{figure}
    \centering
    \includegraphics[scale=0.42]{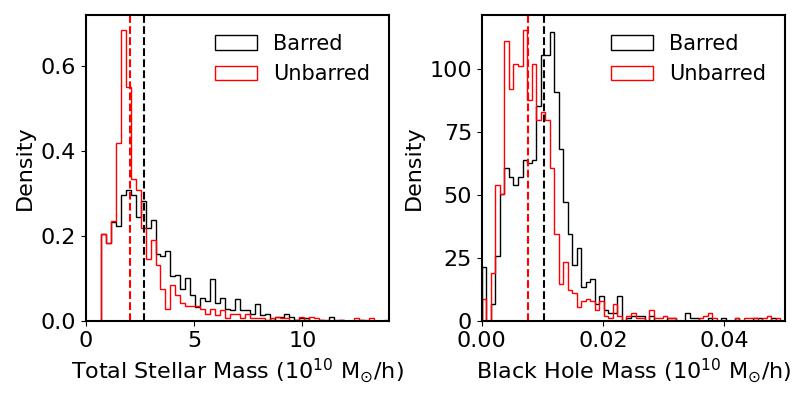}
    \caption{ The left panel shows the distribution of stellar mass for the barred  (black histogram) and unbarred (red histogram) control samples with no duplication. The right panel shows the corresponding black hole mass at z=0 for the barred (black histogram) and unbarred (red histogram) control samples. Black/red vertical lines correspond to the median value of the distribution. }
    \label{fig:total_st_mass_npdup}
\end{figure}
\subsubsection{Sample unbiased using total stellar mass}
{ It is already known that massive stellar disks are prone to bar-type instabilities, and bars in galaxies are preferentially seen in high stellar mass galaxies \citep{Athanassoula.2003}.  The bar fraction, as well as the strength of the bars, are correlated with the stellar mass \citep{Rosas-Guevara.et.al.2020,Diaz2016}. In this context, it is worth checking if the difference in the black hole mass distributions between barred and unbarred galaxies is influenced by the stellar mass parameter. To check this, we constructed a control sample of unbarred galaxies using the stellar mass as the controlled parameter. 

\begin{figure*}
    \centering
    \includegraphics[scale=0.45]{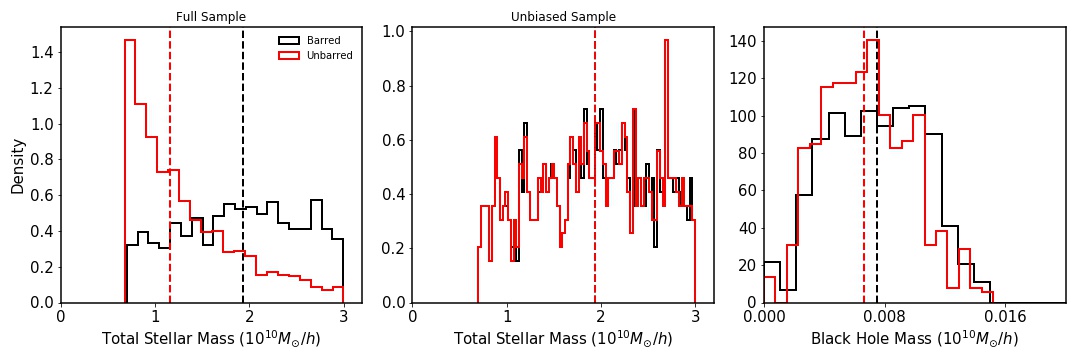}
    \caption{The left panel of the plot depicts the distribution of total stellar mass for both barred and unbarred galaxies, which is truncated at 3 $\times$ $10^{10}$ $M_{\odot}$. The middle panel shows the distribution of the controlled sample for both barred and unbarred galaxies. In the controlled sample, we located the nearest neighbors of similar stellar mass barred galaxies in the full unbarred sample with an upper cutoff. Finally, the right panel of the plot shows the distribution of black hole mass at z=0 in the controlled sample. In all three panels, the vertical dashed lines indicate the median values of the corresponding distributions.}    
    \label{fig:total_st_mass_cut3}
\end{figure*}
Fig.~\ref{fig:full_sample_dist}(b) illustrates the total stellar mass distribution of barred and unbarred galaxies for the full sample, demonstrating that barred galaxies are predominantly located in massive disks compared to unbarred ones. The median stellar mass of barred galaxies is nearly twice that of the unbarred sample. This finding aligns with the observation that bars tend to occur in massive disk galaxies \citep{Rosas-Guevara.et.al.2020}.
 In the controlled sample (left panel of Fig.~\ref{fig:dichotomy_controlled_sample}(b)), the distribution of total stellar mass for barred and unbarred galaxies is similar, unlike the entire sample. The resulting black hole mass distribution of barred and unbarred galaxies is shown in the right panel of Fig.~\ref{fig:dichotomy_controlled_sample}(b). It is clear that the controlled sample does not exhibit the black hole mass dichotomy as seen before in the case of total galaxy mass. 
Given the stark difference in the distribution of stellar mass of barred and unbarred galaxies in the full sample, there arises a valid concern that the generation of the control sample might not be entirely accurate. %We further tested whether stellar mass can be a feasible controlling parameter.
%The difference in the distribution of stellar mass between the two samples could be attributed to the fact that barred galaxies tend to be more prevalent in high stellar mass systems compared to unbarred galaxies. However,
When using the KNN algorithm to create a control sample with similar stellar mass distributions, we found that the same unbarred galaxies were identified as nearest neighbors for barred galaxies with stellar masses greater than $3\times10^{10}M_{\odot}$. The repeated inclusion of the same galaxy in the control sample leads to a situation where this particular galaxy holds a dominating influence over the statistical analysis, which is far from ideal. To address this issue, we modified our algorithm to prevent duplicates in the control sample by not allowing the same unbarred galaxy to be the nearest neighbor of more than one barred galaxy.  This modification did not yield a control sample of unbarred galaxies with an entirely identical stellar mass distribution to that of barred galaxies. The left panel of Fig.~\ref{fig:total_st_mass_npdup} shows the distribution of stellar mass for the barred  galaxies (black histogram) and unbarred (red histogram) control sample galaxies with no duplication. Even in cases where duplicates were omitted, the KNN algorithm identified a control sample of unbarred galaxies generally exhibiting lower masses in comparison to the barred galaxies. This result is understandable, as the KNN algorithm prioritizes the selection of the nearest galaxy based on stellar mass. In situations where nearby galaxies are lacking, the algorithm naturally extends its search to lower mass values in order to identify suitable counterparts.
Due to the aforementioned difference in the stellar mass distribution, we find a marginal dichotomy in the black hole mass distribution between barred and unbarred galaxies when no duplicates in control sample were allowed. The right panel of Fig.~\ref{fig:total_st_mass_npdup} shows the corresponding black hole mass at z=0 for the barred (black histogram) and unbarred (red histogram) control samples. It is evident that the most massive barred galaxies significantly influence the black hole mass distribution in these galaxies.  We also note here that we conducted the analysis without duplicates for all other control parameters as well. However, we observed that this modification didn't impact any of the results, mainly because the initial distributions of the control parameters were not significantly dissimilar to begin with.

In addition to modifying the algorithm to not include duplicates, we also tried creating a control sample with an upper limit to the stellar mass to avoid the high stellar mass galaxies where the KNN algorithm failed to create an ideal control sample dataset.  The left panel of Figure~\ref{fig:total_st_mass_cut3} illustrates the distribution of total stellar mass for the full sample of barred and unbarred galaxies, with an upper mass limit set at $3 \times 10^{10} M_{\odot}$. It is evident that the total stellar mass distribution of unbarred galaxies is skewed towards lower masses, with a median mass of $1.23 \times 10^{10} M_{\odot}$.  Conversely, the distribution for barred galaxies is tilted towards higher values, with a median mass of $2.7 \times 10^{10} M_{\odot}$. Subsequently, we employed the KNN algorithm to identify nearest neighbors within this pruned dataset. %However, when using the KNN algorithm to control the samples, there may not be a candidate unbarred galaxy in the high total stellar mass region corresponding to an individual barred galaxy. To address this issue, we have applied an uppercut of $3 \times 10^{10} M_{\odot}$ to both the barred and unbarred galaxies and
%We then used the KNN algorithm to look for nearest neighbors in the new sample. %The left panel of Figure~\ref{fig:total_st_mass_cut3} shows the distribution of stellar mass for both barred and unbarred galaxies with this uppercut applied. 
The middle panel of Figure~\ref{fig:total_st_mass_cut3} shows the distribution of stellar mass of the controlled sample of barred and unbarred galaxies.
The right panel of Figure~\ref{fig:total_st_mass_cut3} shows the distribution of black hole mass at z=0 for the unbiased sample of barred and unbarred galaxies with an uppercut of $3 \times 10^{10} M_{\odot}$ on the total stellar mass. The distribution of black hole masses at z=0 remains consistent between barred and unbarred galaxies, even within the pruned dataset. In conclusion, there is no discernible dichotomy in the distribution of black hole masses between barred and unbarred galaxies. Notably, the absence of such a dichotomy when controlling for stellar mass implies that stellar mass might indeed be a driver for the observed black hole mass dichotomy. %The black hole mass distribution at z=0 is similar for barred and unbarred galaxies even in the pruned dataset. In summary, we did not observe any dichotomy in the black hole mass distributions between barred and unbarred galaxies. The fact that we did not observe a dichotomy in the black hole mass distributions when unbiased with stellar mass could suggest that stellar mass is a driver for the black hole mass dichotomy. }

%In addition to modifying the algorithm, we also tried creating a control sample with an upper limit to the stellar mass to avoid the high stellar mass galaxies where the KNN algorithm failed. This approach performed better in creating the control sample, and we still did not observe any dichotomy in the black hole mass distributions between barred and unbarred galaxies. The fact that we did not observe a dichotomy in the black hole mass distributions when unbiased with stellar mass could suggest that stellar mass is a driver for the black hole mass dichotomy. 

Previous studies have demonstrated a strong correlation between bars in galaxies and the total stellar mass of the galaxy. The strength of the bars is directly proportional to the stellar mass. It is, therefore, reasonable to assume that black hole feeding would also be more efficient in strongly barred galaxies. If we exclude high-stellar-mass barred galaxies from the sample, we will inadvertently remove galaxies with efficient black hole feeding from the analysis. In such a scenario, we would not expect to observe any significant difference in the black hole mass distributions of barred and unbarred galaxies.
}

\subsubsection{Sample unbiased using total gas mass} 

{ Gas content plays an important role in the evolution of disk galaxies. Bars are thought to drive the gas towards the central regions due to their non-axisymmetric gravitational potential and facilitate the black hole feeding. In turn, the presence of gas also can strengthen or weaken the bar structure as the amount of angular momentum redistribution depends on the material with which angular momentum is exchanged \citep{Athanassoula.2003,Athanassola.et.al.2013}. Many observational studies have investigated the dependence of bar presence and gas content. \citet{Davoust2004} and \citet{Masters2012} reported that the bar fraction is significantly lower among gas-rich disk galaxies than in gas-poor ones. \citet{Zhou2021} reported an increasing trend in the physical size of the bar with an increase in stellar mass and a decrease in gas fraction.
Given that massive gas disks suppress the bar-type instabilities, it is important to check the influence of gas mass on the connection between bars and black hole growth. For this, we unbiased the unbarred galaxy sample using total gas mass as the controlled parameter. Fig.~\ref{fig:full_sample_dist}(c) shows the gas distribution in the full sample, which indicates that the median value of gas mass is marginally higher in unbarred galaxies compared to barred ones. This is in agreement with the previous studies, which found low gas fractions in barred galaxies. The left panel of Fig.~\ref{fig:dichotomy_controlled_sample}(c) shows a similar distribution of total gas mass for the barred and unbarred sample after the controlling exercise. The right panel of Fig.~\ref{fig:dichotomy_controlled_sample}(c) shows the corresponding black hole masses, and it is evident that the dichotomy in the black hole mass distribution between the two galaxy samples persists even after unbiasing with the total gas mass parameter.   The median value of black hole mass distribution peaks at 0.47 $\times$ 10$^8$ M$_\odot$ for unbarred and 1.02 $\times$ 10$^8$ M$_\odot$ for barred galaxies. 
}
%we explored the growth of black hole mass with similar gas content in barred and unbarred galaxies. In this section, we have used the total gas mass as the controlled parameter to obtain the unbiased unbarred galaxy sample. Fig.~\ref{fig:full_sample_dist}(c) shows the gas distribution in the full sample, which indicates that the median value of gas mass is  marginally higher in unbarred galaxies compared to barred ones.  This is in agreement with the previous studies which found that gas fraction is low in barred galaxies. The left panel of Fig.~\ref{fig:dichotomy_controlled_sample}(c) shows that the distribution of barred and unbarred galaxies are pretty similar, given sample is unbiased. Further, we also notice the dichotomy in the distribution of black hole masses in controlled barred and unbarred samples, as seen in the right panel of Fig.~\ref{fig:dichotomy_controlled_sample}(c). The median value of black hole mass distribution peaks at 0.47 $\times$  10$^8$ for unbarred and 1.02 x $10^8$ $M_\odot$ for unbarred galaxies. 

\subsubsection{Sample unbiased using the ratio of total stellar to total halo mass} 
{It is well known that massive disks support the formation of bars. { Similarly, many simulations studied the role of dark matter component on bar formation \citep[for e.g., see,][]{Ostriker.Peebles.1973, Athanassoula.2003, Dubbinski2009}.} While earlier studies with static halos showed that the gravity of the halo suppresses bar formation \citep{Ostriker.Peebles.1973}, recent simulations with live halos reported that the angular momentum exchange enables the bar to grow stronger and longer \citep{Athanassoula2002}. Moreover, the halo concentration,  total halo mass, and spin of the dark matter halo are reported to have an influence on the bar formation \citep{debasitta1998,Athanassoula.2003,saha_Naab2013,Collier2018}.  In this scenario, it is interesting to check how the ratio of stellar mass to halo mass affects the dichotomy of black hole mass distributions presented earlier. Fig.~\ref{fig:full_sample_dist}(d)
shows the distribution of the total stellar to total halo mass ratio for the full sample of barred and unbarred galaxies. A small difference in the distribution is evident from the median of the distributions marked by vertical lines. Here again, we unbiased the unbarred sample using the total stellar to total halo mass ratio, and the resulting distributions are shown in Fig.~\ref{fig:dichotomy_controlled_sample}(d). As seen in the left panel of Fig.~\ref{fig:dichotomy_controlled_sample}(d), the distribution of the ratio of total stellar mass and total halo mass is similar for barred and unbarred galaxies when the sample is controlled. Despite this similarity, we find that the dichotomy in the distribution of black hole masses persists even in the unbiased sample of the barred and unbarred galaxies, as seen in the right panel of Fig.~\ref{fig:dichotomy_controlled_sample}(d).

}
%However It is interesting to see the black hole growth in the sample having a similar ratio of total stellar mass and dark matter halo mass. The distribution of the total stellar to total halo mass ratio is shown in Fig.~\ref{fig:full_sample_dist}(d) for all barred and unbarred sample galaxies. In this section, we have obtained the unbiased sample using the total stellar ratio to halo mass ratio. We find that the distribution of the ratio of total stellar mass and total halo mass is similar for barred and unbarred galaxies when the sample is controlled, as seen in the left panel of Fig.~\ref{fig:dichotomy_controlled_sample}(d). Further, we find that the dichotomy exists in the distribution of black hole masses in the unbiased sample of the barred and unbarred galaxies as seen in the right panel of Fig.~\ref{fig:dichotomy_controlled_sample}(d). This result is similar when we used total galaxy mass (closer to the total halo mass of the galaxy) as a controlling parameter, unlike total stellar mass.  

\subsubsection{Sample unbiased using the ratio of total gas mass to total stellar mass} As discussed in previous sections, the stellar and gas mass have opposite effects on the growth of bar formation. It would be interesting to check the distribution of black hole mass in the two samples with similar  gas to stellar mass ratios. Fig.~\ref{fig:full_sample_dist}(e) shows the distribution of the total gas mass to total stellar mass for the full sample of barred and unbarred galaxies. The medians of the distributions show marked differences between the barred and unbarred samples. So, we obtained the unbiased sample using the ratio of total gas to total stellar mass as the control parameter. We find that the distribution of the ratio of total gas mass and total stellar mass is similar for barred and unbarred galaxies when the sample is unbiased (left panel of Fig.~\ref{fig:dichotomy_controlled_sample}(e)).
Even in this case, we find that the dichotomy exists in the distribution of black hole masses in the unbiased sample of the barred and unbarred galaxies, as shown in the right panel of Fig.~\ref{fig:dichotomy_controlled_sample}(e). { The median value of black hole mass distribution peaks at 0.55 $\times$ $10^8$ $M_\odot$ for unbarred and 1.02 $\times$ $10^8$ $M_\odot$ for barred galaxies.} This result is similar to the case where we used total gas mass as a controlling parameter, but differ from the case when total stellar mass was used as the controlling parameter.

\subsubsection{Sample unbiased using the sum of total stellar mass and total gas mass}
{ As the central black hole can feed on both stars and gas, the difference in the amount of total baryonic mass (stellar mass + gas mass) available in the galaxy can drive the difference in the distributions of black hole mass between barred and unbarred galaxies. To probe if the total baryonic mass has any effect on the black hole mass growth, we repeated the unbiasing exercise for the two samples with the total baryonic mass as the controlled parameter. 
 We find that the distribution of the total baryonic mass is similar for barred and unbarred galaxies when the sample is controlled, as seen in the left panel of Fig.~\ref{fig:dichotomy_controlled_sample}(f). Here again, we find that the dichotomy persists in the distribution of black hole masses even in the unbiased sample of the barred and unbarred galaxies, as shown in the right panel of Fig.~\ref{fig:dichotomy_controlled_sample}(f).  
}
\subsubsection{Summary of the unbiasing exercises}
{ In this section, we carried out unbiasing the galaxy samples using total galaxy mass, total stellar mass, total gas mass, star-to-halo mass ratio, gas-to-star mass ratio, and total baryonic mass as the control parameters. If any of these parameters drive the black hole mass dichotomy, we should have seen similar distributions of black hole mass for the unbiased barred and unbarred galaxies. Except for the stellar mass, we see that the black hole mass dichotomy is preserved even after the unbiasing exercise. 
In the case of stellar mass as the controlled parameter, the distribution of black hole mass at z=0 is similar for both barred and unbarred galaxies. %On further probing, we attribute this disappearance of black hole mass dichotomy to the striking difference in the distribution of stellar mass for the barred and unbarred galaxies which in turn limits the creation of unbiased samples. The significant difference in the stellar mass between the two galaxy samples is expected as bars are known to be preferentially formed in galaxies with massive stellar disks. However, we did not face any issues while creating unbiased samples with other parameters although bars formation is known to be influenced by the gas and halo components. 
We attribute the vanishing of the black hole mass dichotomy after controlling for total stellar mass to the significant influence of stellar mass in the formation of bars and the subsequent efficient fueling of the black holes, allowing for their growth. }
\subsection{Effect of surrounding gas density of galaxy and mergers}
In addition to bars, environmental factors such as galaxy mergers can also contribute to the inflow of gas toward the central regions of a galaxy, thereby facilitating black hole growth. To accurately assess the impact of galactic bars on black hole growth, it is crucial to distinguish the effects of mergers from those of bars \citep{Kormendy.Kennicutt.2004}. In this section, we explore how non-secular evolution events like galaxy mergers affect our analysis. 

\subsubsection{Effect of surrounding gas density}
{We  first investigated the impact of  gas density around the host galaxy on the distribution of black hole mass of barred and unbarred galaxies. We  eliminated any bias in our sample by controlling the surrounding host galaxy gas density as the control parameter. The median black hole mass in barred galaxies peaks higher than in unbarred galaxies, resulting in the black hole mass dichotomy.

{ Additionally, we refined our analysis by considering two additional parameters simultaneously that could influence black hole mass growth: the total mass  and the surrounding neighborhood gas density around the black hole in the host galaxy. To gauge the surrounding gas density, we employ the "BH\_Density" parameter derived from the TNG100 simulations. This parameter represents the local comoving gas density, averaged across the BH's closest neighbors. It pertains to the density within a comoving radius of the sphere encompassing 256 ±4 nearest-neighbor gas cells around the BH. For the galaxies in our sample, this measurement spans from 1 kpc to 56 kpc and necessarily reflect the availability of gas extending beyond the length scales of  the bars.}

To achieve this, we utilized a 2D KNN algorithm that uses the total galaxy mass and the gas density in the vicinity of the black hole as anchor points. This allowed us to identify the nearest neighbors of the unbarred sample galaxies to those of the barred sample, in terms of these two parameters.

Fig.~\ref{fig:2d_knn} illustrates the distribution of black hole masses for the unbiased barred and unbarred samples obtained using this method. We can observe that the median black hole mass in the unbarred sample (0.6 x $10^8$ $M_\odot$) is lower than that in the barred sample (1.01 x $10^8$ $M_\odot$), which confirms the black hole mass dichotomy observed in our earlier results.
 }
\begin{figure}
    \centering
    \includegraphics[scale=0.325]{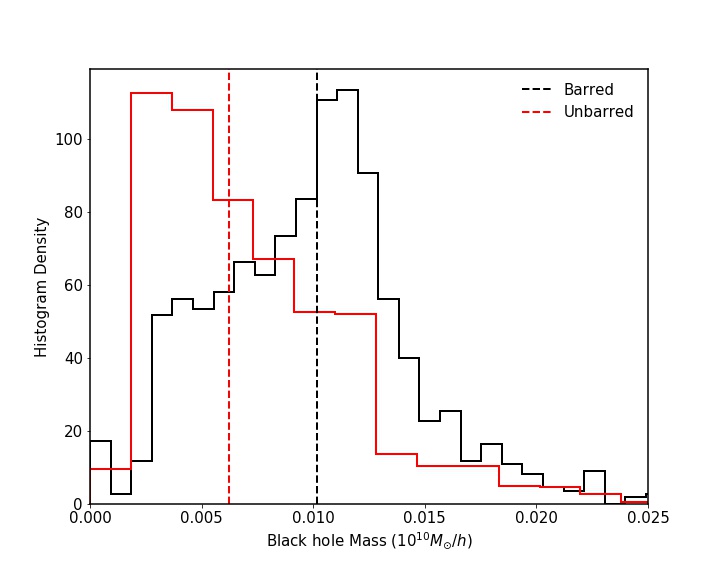}
    \caption{The distribution of black hole mass at z=0 of barred and unbarred samples is obtained using the 2D KNN method by choosing the total galaxy mass and surrounding gas density of the black hole host halo. Black/red vertical lines correspond to the median value of the distribution.}
    \label{fig:2d_knn}
\end{figure}

%\subsubsection{Effect of mergers on black hole growth}

\begin{figure*}
    \centering
\subfigure[Barred Galaxies]{\includegraphics[width=.48\linewidth]{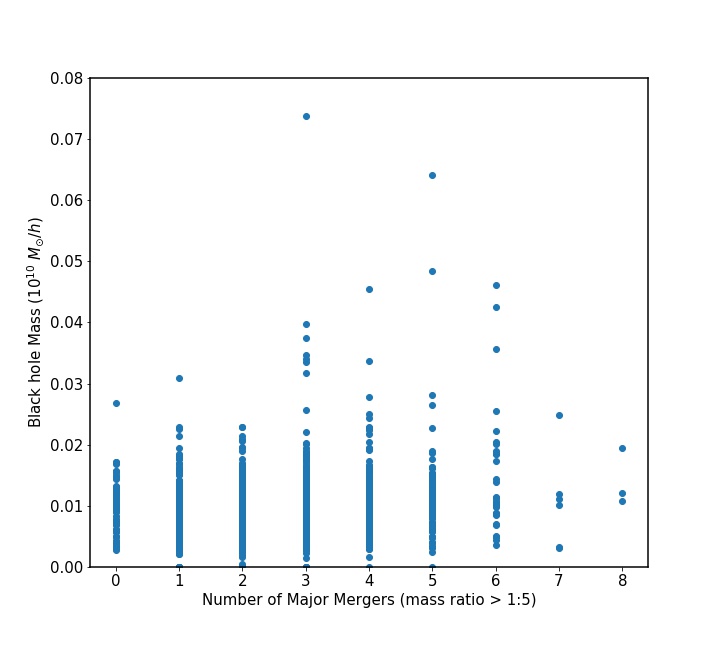}\label{fig:Mergers_barred}  }  
\subfigure[Unbarred Galaxies]{\includegraphics[width=.48\linewidth]{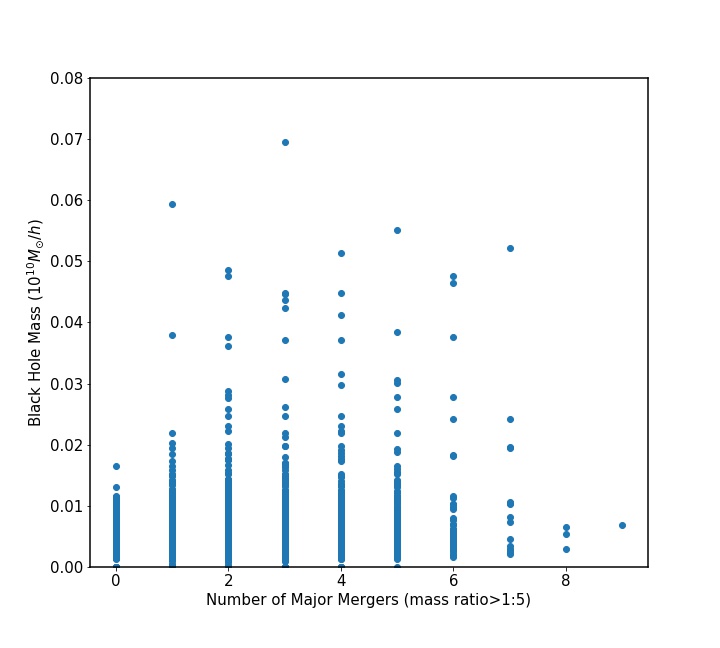} \label{fig:Mergers_unbarred} }
\caption{Scatter plot for the number of major mergers with the black hole masses at z=0 for barred (left panel) and unbarred galaxies (right panel).}
\label{fig:major_mergers}

\end{figure*}

\begin{figure}
    \centering
    \includegraphics[scale=0.325]{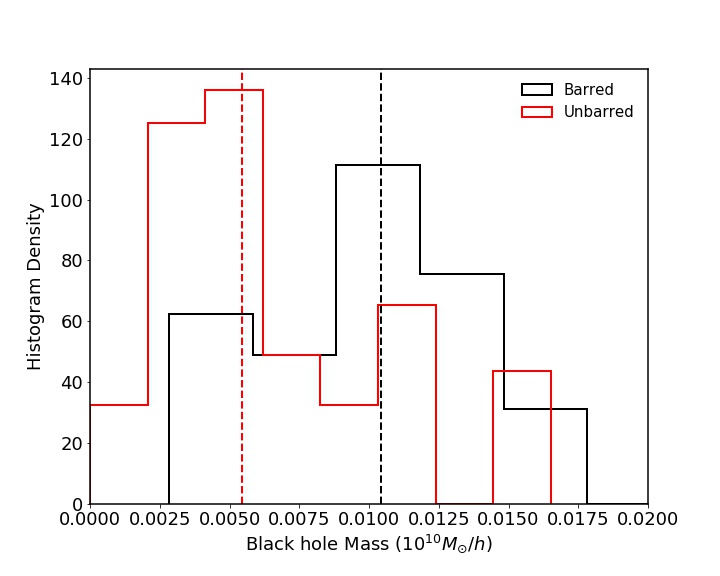}
    \caption{Distribution of black hole mass at z=0 for barred (black histogram) and unbarred (red histogram) galaxies without any merger event during their evolution. Black/red vertical lines correspond to the corresponding median values of the distribution.}
    \label{fig:Mass_dichotomy_no_merger}
\end{figure}

\subsubsection{Galaxies without major mergers} 
{ To disentangle the effect of galaxy mergers, we examined the black hole mass distribution in individual subhalos hosting galaxy disks and constructed a merger tree at the subhalo level using the SubLink algorithm \citep{Gomez.et.al.2015}. By tracing the evolutionary history of these galaxies and excluding those that have undergone major or minor mergers, we can better understand the specific role of bars in black hole mass growth.
}

{We define the major merger as an event for the parent galaxy (subhalo) when a satellite galaxy of mass greater than 1/5 of the parent galaxy falls inside the parent galaxy.  In Fig.~\ref{fig:Mergers_barred}, we present a scatter plot showing the number of major mergers that barred galaxies in our sample have experienced since their formation, plotted against their black hole masses at z=0.  Fig.~\ref{fig:Mergers_unbarred} shows a similar scatter plot for unbarred galaxies. { We notice a tendency for the maximum black hole mass to rise as the number of major mergers experienced by galaxies increases. However, it's important to note that this trend doesn't hold true for galaxies with more than six major mergers. } We also notice from Fig.~\ref{fig:major_mergers} that none of the galaxies in our sample have more than eight major merger events throughout their evolutionary history. Also, only a few galaxies have gone through more than five mergers in our sample. 

To investigate the role of bars in the growth of black holes, we specifically selected galaxies that have not undergone any major merger events. This resulted in 75 barred galaxies and 89 unbarred galaxies. In Fig.~\ref{fig:Mass_dichotomy_no_merger}, we present the distribution of black hole masses for this subset of galaxies. Our findings reveal that even in galaxies that have not experienced any major mergers throughout their evolutionary history, the distribution of black hole masses in barred galaxies is higher than that in unbarred galaxies. This suggests that the presence of a bar may enhance the growth of black holes, strengthening the claim that bars play a significant role in the growth of black holes in galaxies.
}
\subsubsection{Galaxies with small number of  minor mergers}
{
\citet{Rosas-Guevara.et.al.2020}, noted that strongly barred galaxies tend to experience a higher frequency of minor mergers compared to major merger events.  We define a minor merger event as one where the ratio of infalling galaxies is between 1:10 to 1:100. In Fig.~\ref{fig:barred_minor}, we present a scatter plot showing the black hole masses of galaxies plotted against the number of minor merger events they have undergone throughout their evolutionary history.  Our results indicate that the total number of minor mergers is much higher than major mergers, and there is a clear correlation between the number of minor merger events and the black hole mass of host galaxies, similar to the trend observed for major mergers. {  The tendency for  black holes to exhibit greater mass in galaxies that have experienced a higher frequency of mergers, including both minor and major mergers, can be attributed to the fact that these galaxies have undergone more substantial growth, resulting in the proportional growth of their black hole masses.} Fig.~\ref{fig:unbarred_minor} shows the number of minor merger events for an unbarred sample identical to the barred galaxies mentioned above. We can also see a high frequency of minor merger events compared to major mergers and the correlation of increasing black hole mass with an increase in minor merger events. { It is evident that a higher number of mergers signifies a more advanced stage of galactic evolution, leading to an increase in mass.}

To eliminate the effect of minor mergers on black hole growth, we only consider barred and unbarred galaxies that have undergone less than five minor mergers. The sample size becomes too small when we consider galaxies with no minor mergers at all. Fig.~\ref{fig:dichotomy_nominor_merger} shows the distribution of black hole masses for the selected barred and unbarred galaxies. We observe that the dichotomy in black hole masses between barred and unbarred galaxies remains significant even after controlling for the effect of minor mergers. This finding provides further evidence to support the notion that bars play a significant role in enhancing black hole growth by efficiently funneling gas toward the central region of the galaxy.
}
\begin{figure*}
\centering
 \subfigure[Barred Galaxies]{\includegraphics[width=.48\linewidth]{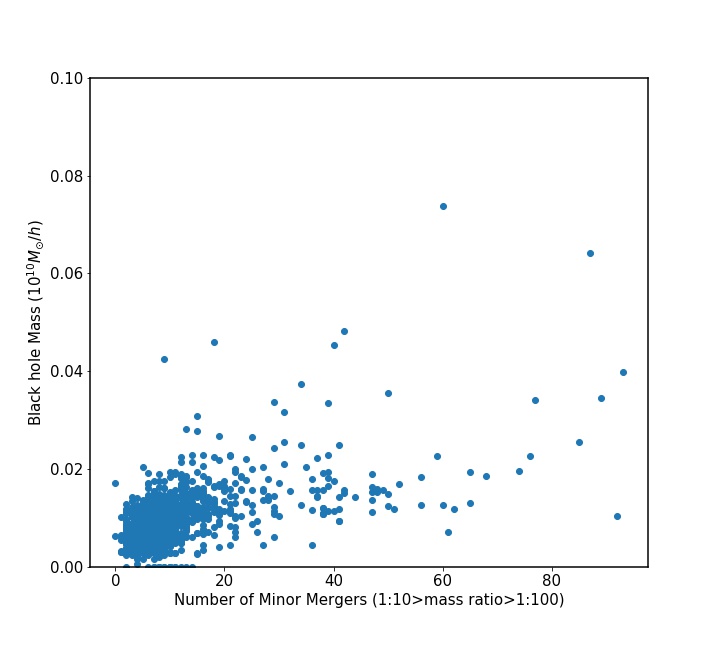} 
 \label{fig:barred_minor}}
\subfigure[Unbarred Galaxies]{\includegraphics[width=.48\linewidth]{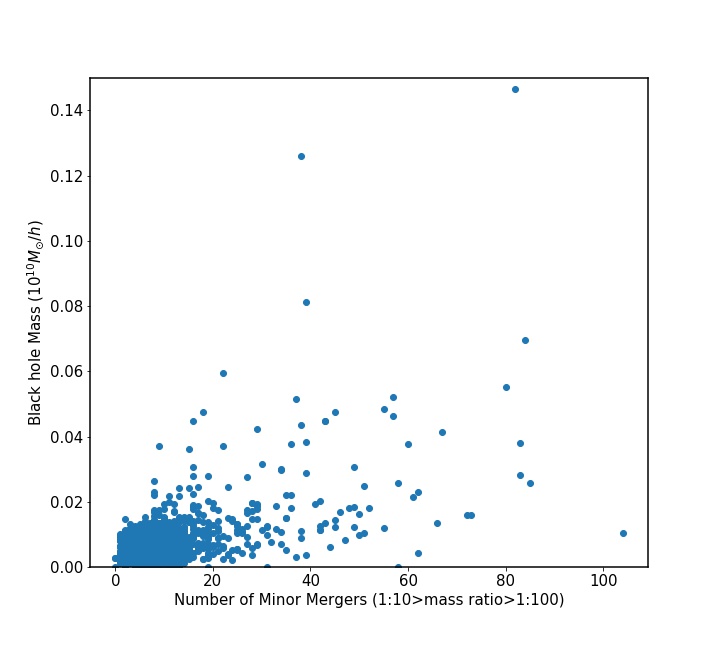} \label{fig:unbarred_minor}}

\caption{The scatter plot of black hole mass at z=0 and the number of minor mergers having mass ratios between 1:10 and 1:100 for barred (black histogram) and unbarred galaxies (red histogram). Black/red vertical lines correspond to the corresponding median values of the distribution. }
\label{fig:minor_mergers}
\end{figure*}

\begin{figure}
    \centering
    \includegraphics[scale=0.325]{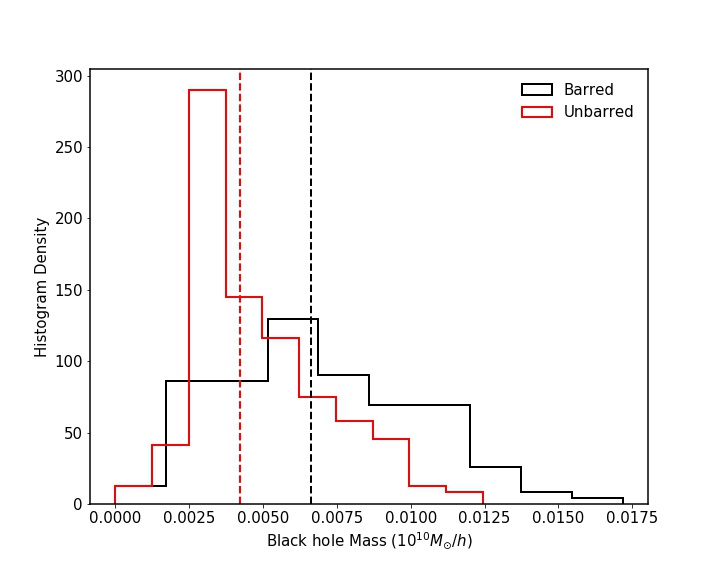}
    \caption{The distribution of black hole masses at z=0 for barred (black histogram) and unbarred (red histogram) galaxies that have not gone through more than five minor mergers. Black/red vertical lines correspond to the corresponding median values of the distribution.}
    \label{fig:dichotomy_nominor_merger}
\end{figure}

\subsection{The black hole mass dichotomy with weak and strong bars}
{
We differentiated strongly and weakly barred galaxies based on the Fourier mode strength cutoff in \citep{Rosas-Guevara.et.al.2020} and ellipticity in \cite{Zhao.et.al.2020} as explained in section \ref{selection_criteria}. Our sample contains 836 strongly barred galaxies and 357 weakly barred galaxies. To investigate the impact of bars on black hole growth, we examined the distribution of black hole masses separately for unbarred, weakly barred, and strongly barred galaxies. Fig.~\ref{fig:WB_SB_BH_Mass} displays the distribution of black hole masses at  z=0 for each galaxy type. The plot indicates that the peak of black hole mass distribution in weakly and strongly barred galaxies is higher compared to unbarred galaxies. This is possible if  bars facilitate black hole mass growth in galaxies. Additionally, Fig.~\ref{fig:WB_SB_BH_Mass} suggests that the growth of black holes in strongly barred galaxies is comparable to that of weakly barred galaxies.  This observation may suggest that although strongly barred galaxies may have a greater availability of gas being funneled towards their central regions, the strong feedback in these galaxies may regulate black hole growth to a similar level as in weakly barred galaxies.
}

\begin{figure}
    \centering
    \includegraphics[scale=0.325]{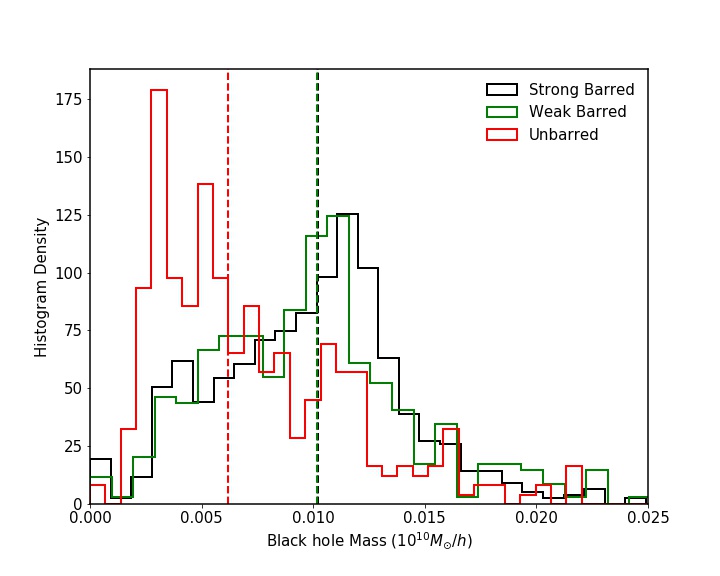}
    \caption{Distributions of black hole mass at z=0  for unbarred (red histogram), weakly barred (green histogram), and strongly barred (black histogram) galaxies. Black/green/red vertical lines correspond to the corresponding median values of the distribution.}
    \label{fig:WB_SB_BH_Mass}    
\end{figure}
\subsection{Time evolution of bar strength and black hole mass}
{ A potential concern in our current analysis is that barred galaxies are selected based on their morphology at z=0. However, bars evolve along with their host galaxy. They form, grow in length and strength, and weaken, sometimes vanishing altogether, either definitively or temporarily. The existence of a robust bar at z=0 does not conclusively indicate that the same bar has persisted from its initial formation. Hence,  it may not be correct to seek a relationship between the final (z=0) black hole mass and the final (z=0) status of the bar. It may be more appropriate to compare the instantaneous black hole mass and the co-evolving bar strength. To address this concern,  we first identified the bar-forming epochs within a sub-sample of galaxies in our dataset. This sub-sample is primarily drawn from the \citet{Rosas-Guevara.et.al.2020} study as the time-evolution of bar strengths is readily accessible from this study. The sub-sample contains 107 and 139 galaxies, which are barred and unbarred, respectively, at z=0. We then control the unbarred galaxies by the total galaxy mass at the bar-forming epoch. We identified a galaxy from the 139  unbarred galaxy sample with a similar total galaxy mass at the bar-forming epoch for each barred galaxy. Following \citet{Rosas-Guevara.et.al.2020}, we defined t$_{norm}$ as the normalized time since the bar formation time, i.e., t$_{norm}$ = (t$_{bar}$ - t$_{lookback}$)/t$_{bar}$ where t$_{bar}$ is the bar age (defined above) and t$_{lookback}$ the look-back time. t$_{norm}$ = 0 corresponds to t$_{lookback}$ = t$_{bar}$, whereas t$_{norm}$ = 1 corresponds to z = 0 (t$_{lookback}$ = 0).  Fig.~\ref{fig:median_bar_bh_mass} shows the median of the black hole mass (solid lines) and bar strength (dashed lines) evolution as a function of t$_{norm}$. It is clear from the figure that once the bar forms, there is little evolution in its strength beyond t$_{norm}$ $\sim$ 0.2. It is encouraging to note that the bar strength never crosses the threshold of 0.2 for unbarred galaxies. The difference between the black hole mass between barred and unbarred galaxies was also maximum at t$_{norm}$ $\sim$ 0.2 when the bars grew to their full size.  After that, the difference between black hole masses decreases, possibly due to feedback effects regulating black hole accretion in the already massive barred galaxies. It is worth noting that this analysis solely draws upon 107 barred galaxies from the \citet{Rosas-Guevara.et.al.2020} sample, a significantly smaller subset compared to our initial analysis.

While our analysis indicates a general trend of stability in bar strength over time, it is important to note that individual cases may exhibit variations influenced by factors such as mergers, flybys, and instabilities.
} %The reason behind this limited number is the availability of time-evolution data for bar strength which is readily accessible only for the  \citet{Rosas-Guevara.et.al.2020} sample.}

\begin{figure}
    \centering
    \includegraphics[scale=0.325]{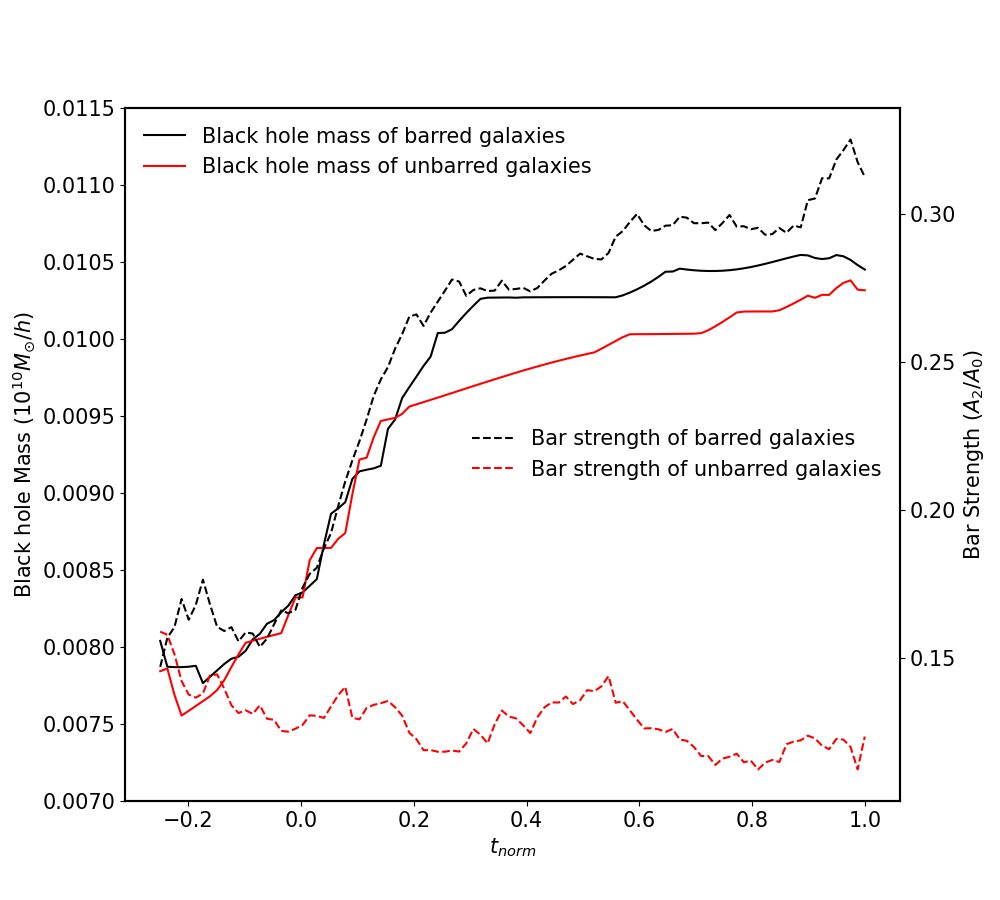}
    \caption{Time
evolution of black hole mass and bar strengths in the barred and unbarred galaxies with time. Solid and dashed lines corresponds to black hole mass and bar strength respectively.  Here t$_{norm}$=0 corresponds to bar forming epoch and t$_{norm}$ =1 corresponds to current epoch. Bar strength is measured as the ratio of m=2 Fourier mode to m=0 Fourier mode of star particles.}
    \label{fig:median_bar_bh_mass}
\end{figure}

\section{Discussion} \label{Discussion}
{
Observational evidence of bar-AGN connection has been argued in various studies where some studies show that barred galaxies preferentially host AGNs \citep{Laine.et.al.2002,Oh.et.al.2012, Galloway.et.al.2015, Alonso.et.al.2018,Silva-Lima.et.al.2022} while other studies claim no connection between AGNs and barred host galaxies \citep{Regan.Mulchaey.1999,Cisternas.et.al.2013,Cheung.et.al.2015,Golding.et.al.2017}. In this study, we have looked for bar-AGN connections in TNG100 disk galaxies.
In previous sections, we demonstrated that black hole masses of barred galaxies  are consistently higher than those of unbarred galaxies in the TNG100 simulations, even after accounting for potential biases. 
This would suggest that the presence of a bar in a galaxy could play a crucial role in facilitating black hole growth by channeling gas to the central regions of the galaxy. In this section, we discuss the implications of our results. 
}

\subsection{Difference in Black hole Seeding}
{
One possibility for the origin of the black hole mass dichotomy observed in barred and unbarred galaxies is that it arises from differences in the seeding of black holes in these systems. In this scenario, barred galaxies may have experienced earlier or more efficient black hole seed formation compared to unbarred galaxies, resulting in larger black hole masses at later times.

As previously mentioned, black holes are seeded in massive halos with a mass exceeding $7.8\times10^{10} M_{\odot}$. To investigate the seeding time of black holes in our analysis, we selected controlled barred and unbarred samples with total galaxy mass as the control parameter and traced the redshift evolution of black hole counts in these sample galaxies. Fig.~\ref{fig:number_of_blackhole} displays the black hole counts in the controlled barred and unbarred samples from redshift 8 to the present epoch.
The plot indicates that black holes in barred galaxies were seeded earlier than those in unbarred galaxies. The difference in black hole seeding between barred and unbarred galaxies is prominent during the redshift range of 7 to 2. This finding suggests that the halos in which bar-forming disks appeared were more massive and formed earlier. This also explains why the disks of barred galaxies tend to be more massive, as their respective dark matter halos formed earlier and had more time to accrete gas than unbarred galaxies, as seen in Fig.~\ref{fig:full_sample_dist}(b). The question immediately arises whether the early seeding of black holes in barred galaxies leads to the observed dichotomy in black hole masses, given the extended period of gas accretion in barred galaxies.  To investigate this, we estimated the maximum mass that can be accreted by the black hole due to early seeding alone, from redshift z=7 to z=2.

Fig.~\ref{fig:accretion_rate_redshift} depicts the evolution of the mean accretion rate for the barred and unbarred galaxy samples while controlling for the total galaxy mass. We observe that the mean accretion rate for both samples increases around redshift 4. The difference in the mean accretion rate between barred and unbarred galaxies reaches its maximum value at around redshift 2, after which the mean accretion rates for both samples begin to decrease. The number of black holes triggered as a function of redshift shows that unbarred galaxies lag behind barred galaxies. For instance, by redshift 6, a total of 200 black holes are already triggered in barred galaxies, whereas in unbarred galaxies, this happens around redshift 5. However, by redshift 2, the number of black holes already triggered in barred and unbarred galaxies becomes similar. Assuming that the difference in accretion rates between barred and unbarred galaxies remains constant over time and equals the maximum difference observed at z=2, multiplying the accretion rate difference by the time elapsed between the time when the difference in black hole seeding is prominent (i.e., 7>$z$>2) would give us the maximum possible mass difference between the two samples. The difference between the mean accretion rate of black holes in barred and unbarred galaxies ($\Delta\dot{M}_{BH}$) from z=7 to z=2 is given by $10^6 M_{\odot}/(.978 Gyr)$. The maximum difference in black holes mass of barred and unbarred samples which this early black hole seeding can lead is given by:
\begin{equation}
    \Delta M_{BH} \leq \Delta\dot{M}_{BH}\times(t_{z=7}-t_{z=0}) = 1.4 \times 10^6 M_{\odot}
\end{equation}

\begin{figure}
    \centering
    \includegraphics[scale=.35]{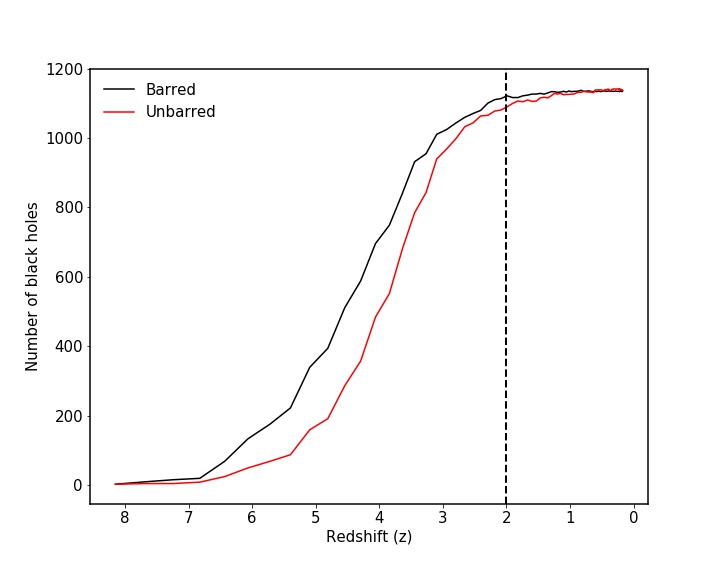}
    \caption{Cumulative distribution of the number of black holes as a function of redshift in the controlled sample of barred (black curve) and unbarred (red curve) galaxies.  The dashed/black vertical line shows the redshift when barred and unbarred samples have an equal number of black holes.  }
    \label{fig:number_of_blackhole}
\end{figure}

\begin{figure}
    \centering
    \includegraphics[scale=0.28]{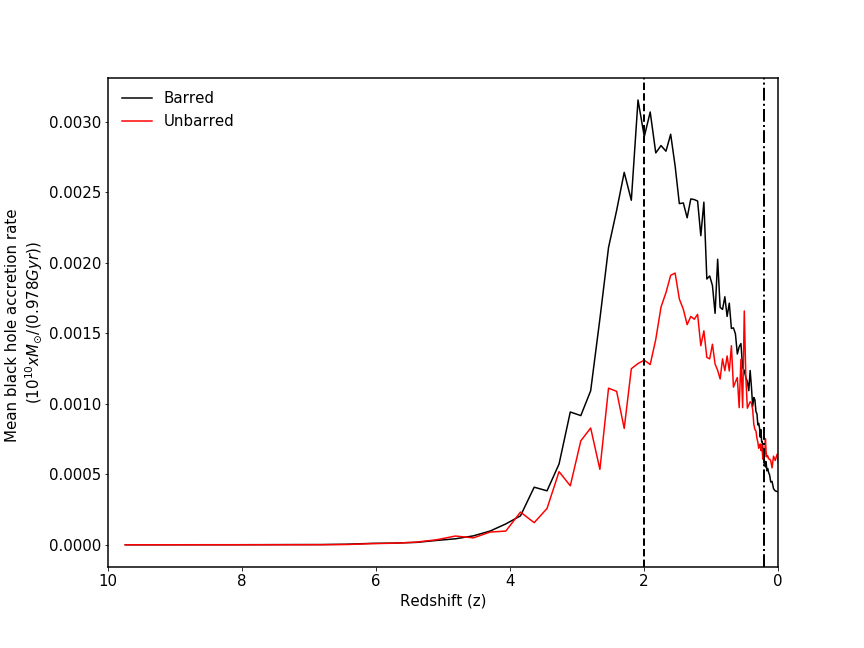}
    \caption{The figure shows the evolution of the mean accretion rate of controlled barred (black curve) and unbarred (red curve) samples with redshift. The dashed vertical line shows the redshift when the difference in mean accretion rate is maximum between barred and unbarred samples. The dot-dashed vertical line indicates the epoch when the mean accretion rate of black exceeds that in unbarred galaxies compared to barred ones.  }
    \label{fig:accretion_rate_redshift}
\end{figure}

\begin{figure}
    \centering
    \includegraphics[scale=0.28]{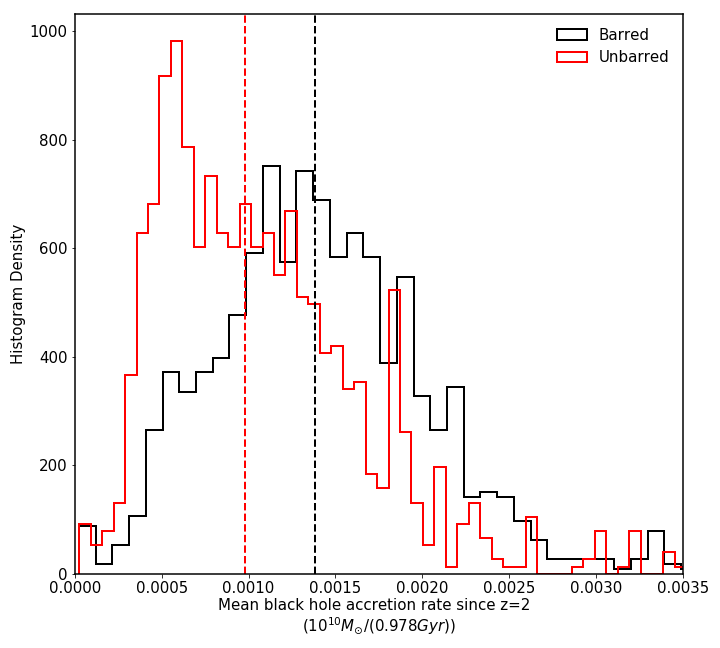}
    \caption{The figure shows the distribution of the average accretion rate of the black holes between 2 $<$ z $<$ 0 for the controlled barred (black histogram) and unbarred  (red histogram) samples. Black/red vertical lines correspond to the corresponding median values of the distribution. }
    \label{fig:meanaccretion_rates_sincez2}
\end{figure}

The calculated value of $\Delta M_{BH}$ resulting from the early seeding effect is approximately one order of magnitude lower than the observed black hole mass difference between barred and unbarred galaxy samples at z=0, which is $4\times10^7 M_{\odot}$. It is worth noting that this calculation represents an upper limit on the potential effect of early seeding. The actual time required for unbarred galaxies to catch up with the number of black holes in barred galaxies is much shorter. Therefore, we can conclude that early seeding of black holes in the barred sample alone is unlikely to be solely responsible for the observed dichotomy between barred and unbarred galaxies in terms of black hole mass at the current epoch.

Figure~\ref{fig:accretion_rate_redshift} also shows that the mean accretion rate of barred galaxies falls below that of unbarred galaxies at around z$\sim$ 0.2. This trend could be interpreted as the result of more significant kinetic feedback in barred galaxies. As discussed in section \ref{simulations}, for black holes with masses greater than $10^{8.5} M_{\odot}$, kinetic feedback is the most effective feedback mechanism, which imparts momentum to the surrounding gas, regulating the accretion rate. The majority of the barred galaxy samples have black hole masses greater than $10^{8.5} M_{\odot}$, whereas only a small fraction of unbarred galaxies have black hole masses above this threshold, as evident from Fig.~\ref{fig:Dichotomy_full_sample}. This suggests that the effects of kinetic feedback began to regulate gas availability for accretion at around redshift 0.2.
}

\subsection{Average accretion rates since bar forming epochs} \label{avg acc rate}
%In Fig.~\ref{fig:accretion_rate_redshift} we see that the barred sample's mean accretion rates are higher than the unbarred sample since z=2 to z=0.2 for which the trend reverses until the current epoch. Why does the presence of a bar after z=0.2 enhance the black hole accretion rate? We want to point out that the kinetic feedback which regulates the gas accretion on a black hole is responsible for this reversing trend of accretion rate. We have already discussed in section \ref{simulations} that for a black hole massive that $10^{8.5} M_{\odot}$ most prominent mode of feedback is kinetic, which imparts momentum to surrounding gas leading to regulation of accretion rate. The black hole mass in most of the barred samples is greater than $10^{8.5} M_{\odot}$ compared to a small fraction on unbarred ones as clearly seen in Fig.~\ref{fig:Dichotomy_full_sample}, which confirms the triggering of efficient feedback mode in the barred sample. 
{
\cite{Rosas-Guevara.et.al.2020} shows that most of the bars in the disks have formed around $z \approx 2$. We expect the bar to channel gas in the central region and enhance the availability of gas in the black hole vicinity \citep{Regan&Teuben.2004}. To investigate the influence of bars on AGN fueling, we calculated the average accretion rate of both the barred and unbarred samples from the formation epoch of bars (z $\sim$ 2) to the present epoch.

Fig.~\ref{fig:meanaccretion_rates_sincez2} shows the average accretion rate of black holes in the barred and unbarred samples from z=2 to the current epoch. The plot indicates that the median value of the accretion rate is higher in barred galaxies than in unbarred galaxies, with a difference between the two medians ($\Delta \dot{M}_{BH}$) of approximately $3\times10^6 M{\odot}/(.978 Gyr)$.  To estimate the difference in black hole mass between the two samples, we multiplied the average accretion rate difference with the elapsed time between 2$<$z$<$0. 
\begin{equation}
    \Delta M_{BH} \approx \Delta \dot{M}_{BH} \times (t_{z=2}-t_{z=0}) = 4 \times 10^7 M_{\odot}
\end{equation}
Using the calculated difference in mean accretion rate, we estimated a mass difference that can account for the observed discrepancy in black hole masses between the barred and unbarred samples at the current epoch. This mass difference is comparable to the observed difference in black hole masses between the two samples (as shown in the right panel of Fig.~\ref{fig:dichotomy_controlled_sample}(a)), providing evidence that the presence of a bar enhances black hole growth by facilitating gas inflow towards the central regions of a galaxy.
}

\subsection{Stellar mass, bar formation and black hole growth} \label{stellar mass discussion}
{
Stellar mass is known to be one of the key factors influencing the formation and strength of bars in galaxies. Observational studies have shown that the presence of a bar is more common in massive galaxies than in low-mass galaxies \citep[for e.g.,][]{Cheung2015}. Also, it has been observationally found that the fraction of galaxies with bars increases with increasing stellar mass \citep[e.g.,][]{Masters2012}. This trend is also supported by numerical simulations. For example, simulations by \citep{Athanassoula.2003} show that bars are more likely to form in galaxies with massive stellar disks. { \citet{Zana2022} conducted a detailed statistical analysis of the bar population in the TNG50 simulation, affirming that the prevalence of bars is greater in galaxies with higher masses.} The prevalence of bars in massive galaxies can be attributed to the fact that the formation of a bar requires a significant concentration of gas and stars in the central region of a galaxy, which is more likely to occur in massive galaxies due to their deeper potential wells. In addition, massive galaxies tend to have more dynamically hot stellar disks, which are less susceptible to fragmentation and, therefore, more likely to form bars \citep{Athanassoula.2003}.
Additionally, \citet{Athanassola.et.al.2013,Rosas-Guevara.et.al.2020} found that the strength and longevity of bars increase with the mass of the galaxy, with more massive galaxies producing stronger and longer-lasting bars. One possible explanation for this relationship is that massive disks are more stable to perturbations, which allows bars to persist for longer periods. This increased stability arises because the gravitational potential of a massive disk can absorb the energy of perturbations more effectively, preventing them from disrupting the bar. 
 Therefore, there is ample evidence to suggest that stellar mass plays a critical role in bar formation in galaxies \citep{Fujii.et.al.2018}. More massive galaxies tend to have larger, more stable disks that are better suited to supporting long-lived bars \citep{Howthorn.et.al.2023}.

In our analysis of comparing the black hole masses of barred and unbarred TNG100 galaxies at z=0, a clear dichotomy is observed in black hole mass distribution, with the barred galaxies exhibiting higher black hole masses.  Despite controlling the samples with various parameters such as galaxy mass and gas mass, the dichotomy in the black hole mass distribution between barred and unbarred galaxies persisted. However, the previously observed dichotomy disappeared after controlling the samples with the total stellar mass parameter. This result suggests that the total stellar mass may be a potential driver for the observed dichotomy in the black hole mass distribution. 

It is possible that galaxy mergers play a major role in the formation and evolution of bars in galaxies. The gravitational interaction and tidal forces from a galaxy merger can transfer angular momentum to the gas and stars in the disk, forming bars.  Numerical simulations have shown that both major and minor mergers can lead to bar formation by perturbing the disk and producing non-axisymmetric features like bars and spirals \citep{Hernquist1995,Cavanagh2020}. However, some studies suggest that bars can form and persist without the need for a merger \citep{debasitta1998,Athanassoula.2003}. { Furthermore, the studies by \citet{Zana2018a, Zana2018b} indicate that external gravitational interactions, such as mergers or flybys, have the potential to impede the evolutionary process of bars within galaxies.} The exact role of mergers in bar formation and evolution is still a topic of active research \citep{ghosh.et.al.2022}. If galaxy mergers were the reason for bar formation, it is possible that the same merger events can channel the gas toward the center enabling black hole growth. If this is the case, the relationship between bars and black hole mass observed in our analysis may be indirect and instead be due to the effect of mergers on both bar formation and black hole growth. To investigate the role of mergers on black hole mass growth, we constructed a sample of barred and unbarred galaxies that have no major mergers / a small number of minor mergers. We find that even in this case, the black hole mass dichotomy exists between barred and unbarred galaxies. This suggests that while mergers may play a role in bar formation and subsequent black hole growth, they are not the sole driving factor for the observed connection between bars and black hole mass.

Bars can transfer the angular momentum within the disk of the galaxy and channel the gas towards the central regions, where it can fuel star formation or accrete onto a central supermassive black hole. Observationally, it has been shown that barred galaxies have a larger molecular gas density in the central regions compared to unbarred galaxies \citep{Sakamoto.et.al.1999}. However, for accretion onto black holes to occur, the gas must be transported to within a few parsecs of the galaxy center, which is closer than the typical location of the Inner Lindblad Resonance of the bar. Additional mechanisms like oscillations of gas flow orbits or bars within bars are invoked to drive further inflow and feed the AGN \citep{Sholsman.et.al.1989}. Many double-barred galaxies have been found through observations. For e.g., \citet{Erwin2004} presented a catalog of 67 double-barred galaxies. Around one-third of barred galaxies are found to host a short inner bar of  radius $<$1 kpc \citep{Erwin2002,Laine.et.al.2002,Erwin2004}. Furthermore, just like the secondary bars, nuclear instabilities like rings and spirals are also thought to facilitate black hole fueling.  However, it is possible that the gas transported by bars is consumed in star formation before it can reach the black hole, thereby limiting the effectiveness of bars in fueling black hole growth. The relative importance of these processes may depend on various factors, such as the gas content and dynamics of the host galaxy. Using the TNG100 simulations, \citet{Rosas-Guevara.et.al.2020} found that strongly barred galaxies have a significantly higher median black hole mass compared to unbarred galaxies, with this difference becoming more pronounced shortly after the epoch of bar formation. While the mean accretion rate of barred galaxies decreases significantly after the formation of the bar, primarily due to the impact of AGN feedback, their analysis clearly suggests that bars have played a crucial role in facilitating the feeding of black holes in these galaxies, as well as enhancing central star formation.

Overall, the connection between stellar mass, bar formation, and black hole growth is complex and multifaceted, with various mechanisms at play, such as gas inflows due to bars, mergers, secular processes, and feedback effects.  Nevertheless, our analysis in this study indicates that bars can play a crucial role in feeding black holes, particularly in galaxies with massive stellar disks.
}

\section{Conclusions} \label{conclusion}
{
 In this article, we analyzed the black hole mass distribution in barred and unbarred galaxies using the TNG100 magneto-hydrodynamic cosmological suite of simulations. We examined 1193 barred and 2738 unbarred galaxy samples while controlling for various parameters. Our main findings are listed below:
\begin{itemize}
    \item There is a clear dichotomy in the distribution of black hole masses in barred and unbarred galaxies. The median black hole mass of the unbarred sample is higher than that of unbarred ones, and the difference between median black hole mass  $\Delta M_{BH} \approx 4\times10^7 M_{\odot}$. 
    \item We removed any underlying bias in our analysis by creating control samples with various parameters like total galaxy mass, total stellar mass, total gas mass, stellar-to-halo mass ratio, gas-to-stellar mass ratio, and total baryonic mass.   Except for the controlling with total stellar mass, we found that the black hole mass dichotomy is reproduced in all other cases. Therefore, we conclude that stellar mass is the primary driver of the black hole mass dichotomy between barred and unbarred galaxies at the current epoch. This can be attributed to the fact that bars require massive stellar disks for their formation, and black hole fuelling is more efficient when the bars are stronger.
    \item  We disentangled the effect of galaxy mergers on black hole growth by removing galaxies with any major/minor mergers from our analysis. The results of this analysis also show a higher mean black hole mass for barred galaxies compared to unbarred galaxies, which further strengthens the role of bars in fueling black hole activity.
    \item The seeding of black holes occurs earlier in barred samples compared to the unbarred sample, in agreement with the earlier studies that bars are formed in massive galaxies. However, this early seeding effect can only lead to a maximum black hole mass difference of $3\times10^6 M_{\odot}$, much lower than the observed black hole dichotomy between barred and unbarred samples at the current epoch. 
    \item The high mean black hole mass in barred galaxies compared to unbarred ones is explained ($\Delta M_{BH} \approx 4 \times 10^7 M_{\odot}$) with the higher mean accretion rate from bar forming epoch redshift z$\approx$2 to z=0.
    \item { Therefore, we suggest a potential association between AGN activity and the presence of a bar within the host galaxy.}

\end{itemize}
 
Future simulations with higher resolution and improved physical modeling can provide a more detailed understanding of the physical mechanisms responsible for the connection between bars and AGN activity, ultimately leading to a more comprehensive understanding of the co-evolution of black holes and their host galaxies.
}

\section*{Acknowledgements}

The Authors would like to thank the Referee for the helpful suggestions that improved the quality of the manuscript.  MV acknowledges support from DST-SERB in the form of core research grant (CRG/2020/1657). 
The IllustrisTNG simulations were undertaken with compute time awarded by the Gauss Centre for Supercomputing (GCS) under GCS Large-Scale Projects GCS-ILLU and GCS-DWAR on the GCS share of the supercomputer Hazel Hen at the High Performance Computing Center Stuttgart (HLRS), as well as on the machines of the Max Planck Computing and Data Facility (MPCDF) in Garching, Germany. The authors thank all the members of the IllustrisTNG team for making the simulation data public.
%%%%%%%%%%%%%%%%%%%%%%%%%%%%%%%%%%%%%%%%%%%%%%%%%%
\section*{Data Availability}

The data is publicly available on IllustrisTNG website as provided by IllustrisTNG team. https://www.tng-project.org/data/



\bibliographystyle{mnras}
\bibliography{example} % if your bibtex file is called example.bib

\bsp	% typesetting comment
\label{lastpage}
\end{document}